\documentclass[%
 reprint,
 amsmath,amssymb,
 aps,
]{revtex4-1}

\usepackage{graphicx}
\usepackage{dcolumn}
\usepackage{bm}

\begin{document}

\preprint{APS/123-QED}

\title{$^{139}$La and $^{63}$Cu NMR investigation of charge order in La$_{2}$CuO$_{4+y}$ ($T_{c}=42$~K)}

\author{T.\ Imai} 
\affiliation{Department of Physics and Astronomy, McMaster University, Hamilton, Ontario L8S4M1, Canada}
\affiliation{Canadian Institute for Advanced Research, Toronto, Ontario M5G1Z8, Canada}
\author{Y.\ S.\ Lee} 
\affiliation{Stanford Institute for Materials and Energy Sciences, Stanford National Accelerator Laboratory, Menlo Park, CA 94025}
\affiliation{Department of Applied Physics, Stanford University, Stanford, CA 94305}

\date{\today}

\begin{abstract}
We report $^{139}$La and $^{63}$Cu NMR investigation of the successive charge order, spin order, and superconducting transitions in super-oxygenated La$_2$CuO$_{4+y}$ single crystal with stage-4 excess oxygen order at $T_{stage}\simeq 290$~K.  We show that the stage-4 order induces tilting of CuO$_6$ octahedra below $T_{stage}$, which in turn causes $^{139}$La NMR line broadening.  The structural distortion continues to develop far below $T_{stage}$, and completes at $T_{charge}\simeq 60$~K, where charge order sets in.  This sequence is reminiscent of the the charge order transition in Nd co-doped La$_{1.88}$Sr$_{0.12}$CuO$_4$ that sets in once the low temperature tetragonal (LTT) phase is established.  We also show that the paramagnetic $^{63}$Cu NMR signals are progressively wiped out  below $T_{charge}$ due to enhanced low frequency spin fluctuations in charge ordered domains, but the residual $^{63}$Cu NMR signals continue to exhibit the characteristics expected for optimally doped superconducting CuO$_2$ planes.  This indicates that charge order in La$_2$CuO$_{4+y}$ does not take place uniformly in space.  In addition, unlike the typical second order magnetic phase transitions, low frequency Cu spin fluctuations as probed by $^{139}$La nuclear spin-lattice relaxation rate do not exhibit critical divergence at $T_{spin}$($\simeq T_{c}$)=42~K.  These findings, including the spatially inhomogeneous nature of the charge ordered state, are qualitatively similar to the case of La$_{1.885}$Sr$_{0.115}$CuO$_4$ [T. Imai et al., Phys. Rev. B {\bf 96} (2017) 224508, and A. Arsenault et al., Phys. Rev. B (submitted)], but both charge and spin order take place more sharply in the present case. 
\end{abstract}

\maketitle


\section{\label{sec:level1} Introduction}
Recent success in detecting the charge order Bragg peaks in many different classes of high $T_c$ cuprates based on modern x-ray scattering techniques \cite{Comin_Review} has led to renewed interest in the charge order in La$_2$CuO$_4$-based cuprates.  In the case of Nd$^{3+}$ co-doped La$_{1.48}$Nd$_{0.4}$Sr$_{0.12}$CuO$_{4}$, in which Tranquada et al. initially discovered the charge order phenomenon \cite{Tranquada}, charge order sets in at $T_{charge} \sim 65$~K only after the structural phase transition from the low temperature orthorhombic (LTO) to the low temperature tetragonal (LTT) phase takes place at  $T_{LTT} \sim 70$~K.  

The LTT structure induced by Nd$^{3+}$ severely suppresses $T_c$ from $\sim 38$~K to $\sim 6$~K.  Accordingly, many researchers speculated early on that charge order is a byproduct of the LTT structure with low $T_c$ and hence absent in the superconducting CuO$_2$ planes, despite our early NMR reports that even superconducting La$_{1.885}$Sr$_{0.115}$CuO$_{4}$ ($T_{c} = 30$~K) with the LTO structure exhibits nearly identical NMR anomalies observed at the charge order transition of La$_{1.48}$Nd$_{0.4}$Sr$_{0.12}$CuO$_{4}$ \cite{HuntPRL, SingerPRB, HuntPRB, ServiceScience}.  Recent confirmation of the existence of charge order Bragg peaks in La$_{1.885}$Sr$_{0.115}$CuO$_{4}$ \cite{Croft, Thampy, He} finally settled the two decades old controversy.  

On the other hand, these latest x-ray scattering work on La$_{1.885}$Sr$_{0.115}$CuO$_{4}$ also revealed that charge order sets in gradually starting from as high as $T_{charge} \simeq 80$~K \cite{Croft, Thampy, He}, indicating that our initial NMR report of $T_{charge} \simeq 50$~K \cite{HuntPRL} overlooked the gradual onset.  Therefore, we recently revisited $^{63}$Cu  \cite{ImaiPRB2017} and $^{139}$La NMR  \cite{ArsenaultPRB2017} evidence for charge order in La$_{1.885}$Sr$_{0.115}$CuO$_{4}$.  We demonstrated that the normal $^{63}$Cu NMR signals is gradually wiped out below $T_{charge} \simeq 80$~K, and the lost spectral weight is transferred to a wing-like anomalous $^{63}$Cu NMR signal with extremely fast NMR relaxation rates.  The transverse $T_2$ relaxation time of the wing-like signal is so short that one can detect it only with the NMR pulse separation time shorter than $\tau \sim 4~\mu$s.  Our NMR results confirmed that charge order does {\it not} proceed uniformly in space in La$_2$CuO$_4$-based cuprates \cite{HuntPRL}, and a diminishing fraction of CuO$_2$ planes remains unaffected by charge order even at $T_{c}=30$~K.  

These recent developments raised a new question.  Why is $T_{charge}\simeq 80$~K in orthorhombic La$_{1.885}$Sr$_{0.115}$CuO$_{4}$ even higher than $T_{charge} = 65$~K of La$_{1.48}$Nd$_{0.4}$Sr$_{0.12}$CuO$_{4}$?  Do the LTT structure and/or disorder induced by Nd$^{3+}$ substitution play a role?  Clearly, we need more investigations to understand the structural effects on charge order.

The latest addition in the growing list of charge ordered cuprates is the super-oxygenated La$_{2}$CuO$_{4+y}$ ($y \sim 0.11$) with $T_{charge} \simeq 60$~K as determined by x-ray diffraction \cite{Wells, He}.   La$_{2}$CuO$_{4+y}$ undergoes simultaneous superconducting and spin density wave (SDW) order at $T_{c} \simeq T_{spin} = 42$~K \cite{LeePRB1999, Savici} within the charge ordered state, similarly to La$_{1.885}$Sr$_{0.115}$CuO$_{4}$ with $T_{c} \simeq T_{spin} = 30$~K \cite{Kimura} and $T_{charge} \simeq 80$~K \cite{Croft, Thampy, He}.  Unlike the La$_{2}$CuO$_{4}$-based materials substituted by Sr$^{2+}$ or Ba$^{2+}$, it is the excess O$^{2-}$ ions located at the interstitial sites that donate holes into CuO$_2$ planes \cite{Jorgensen, Chou, Wells_ZP, Wells_Science, LeePRB1999, KhaykovichPRB2002, KhaykovichPRB2003, JorgePRB2004}.  These O$^{2-}$ ions are highly mobile near room temperature, and form a stage-4 superstructure below $T_{stage} \simeq 290$~K \cite{Wells_ZP, Wells_Science}.  In the stage-4 ordered phase, interstitial O$^{2-}$ ions occupy LaO layers separated by four CuO$_2$ planes, and create anti-phase domain boundaries for the tilting direction of CuO$_6$ octahedra \cite{Wells_Science}.  Upon further cooling, O$^{2-}$ ions lose mobility around $T_{mobility} \simeq 200$~K \cite{Kremer} and develop a superstructure within the plane \cite{, Wells_Science}.   Unlike the quenched disorder induced by Sr$^{2+}$ or Ba$^{2+}$ substitution, super-oxygenated La$_{2}$CuO$_{4+y}$ therefore possesses unique, {\it annealed disorder} \cite{Wells_Science}. 

In this paper, we report  $^{139}$La and $^{63}$Cu NMR study of a La$_{2}$CuO$_{4+y}$ ($y \sim 0.11$) single crystal, taking full advantage of the unique characteristics of NMR as a local probe.  We identify NMR anomalies associated with the stage-4 oxygen order at $T_{stage} \simeq 290$~K and $T_{mobility} \simeq 200$~K.  We show that, at the slow observation time scale of NMR, some $^{139}$La sites remain unaffected by the stage-4 order down to $T_{charge} \simeq 60$~K.  This may be an indication that the stage-4 order plays an analogous role as the LTT structural transition in La$_{1.48}$Nd$_{0.4}$Sr$_{0.12}$CuO$_{4}$ \cite{Tranquada} and La$_{1.88}$Ba$_{0.12}$CuO$_{4}$ \cite{Fujita}, in which charge order does not set in until the LTT structure is established below $T_{LTT}$.  We also show that charge order does not proceed homogeneously in space.  Nearly 1/3 of the volume fraction of the CuO$_2$ planes is still unaffected by charge order at $T_c$, and exhibits NMR properties expected for canonical superconducting CuO$_2$ planes. 

\section{\label{sec:level1} Experimental }

We super-oxygenated a single crystal of La$_{2}$CuO$_{4}$ using the electrochemical doping technique \cite{Chou, LeePRB1999}.  We determined the superconducting critical temperature $T_{c} = 42$~K of our crystal with the Superconducting Quantum Interference Device (SQUID).  Both elastic neutron scattering \cite{LeePRB1999} and $\mu$SR \cite{Savici} measurements found the onset of the spin density wave order at $T_{spin}$ ($= T_{c}$) $= 42$~K for a different piece of crystal cut from the same boule.  Unlike the case of lower excess O$^{2-}$ composition \cite{Hammel}, we did not find evidence for a phase separation into antiferromagnetic and superconducting domains in the $^{139}$La NMR lineshape nor the bulk susceptibility data.

We used a piece of single crystal with the approximate dimensions of $5 \times 5 \times 1$~mm for our NMR measurements.  We conducted all the NMR measurements at M.I.T. between 1997 and 1999 using the NMR spectrometer built with a Tecmag Aries console.  The Aries was a state-of-the-art NMR system at the time, but the spectrometer dead time $t_{dead} = 8$ to 10 $\mu$s caused by the ring down was very long in today's standard.  Accordingly, we conducted all the NMR measurements with $\tau = 12~\mu$s or longer, and hence we were unable to detect the wing-like $^{63}$Cu NMR signals that we recently discovered in the charge ordered state of  La$_{1.885}$Sr$_{0.115}$CuO$_{4}$ \cite{ImaiPRB2017}; the wing-like NMR signal has extremely fast $T_1$ and $T_2$ relaxation rates, and its detection requires much shorter spectrometer dead time ($t_{dead} = 2$ to 4 $\mu$s).

\section{\label{sec:level1} Results and Discussions}
\subsubsection{Stage-4 order as probed by $^{139}$La NMR}

\begin{figure}
\centering
\includegraphics[width=3.2in]{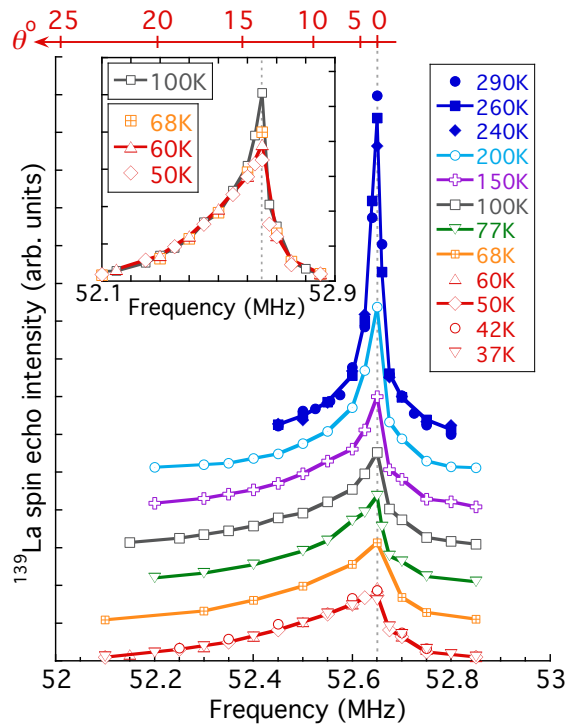}
\caption{Representative $^{139}$La NMR lineshapes in $B_{ext} = 8.7521$~T applied along the c-axis.    For clarity, the vertical origin is shifted at different temperatures.  The gray vertical dashed line marks the peak frequency $^{139}f_{o} = 52.65$~MHz.  Asymmetrical line broadening becomes significant below $\sim 220$~K.  (Inset) The very sharp, needle shaped peak at $^{139}f_{o} = 52.65$~MHz, which arises from $^{139}$La sites neighboring to untilted CuO$_6$ octahedra, remains observable down to $\sim 60$~K.  
}\label{cs}
\end{figure}

\begin{figure}
\centering
\includegraphics[width=3.6in]{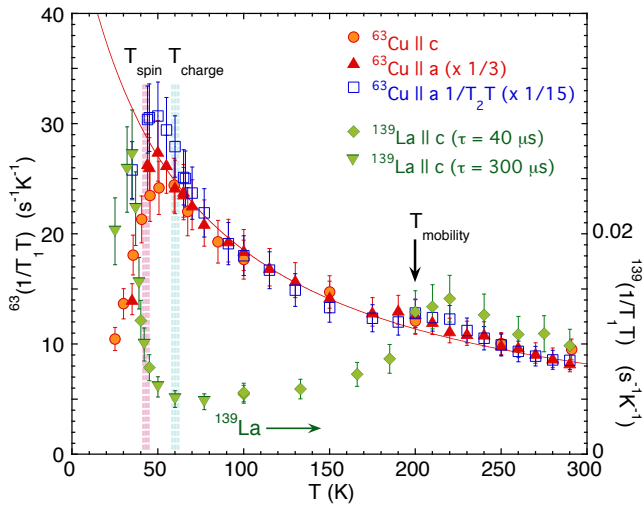}
\caption{Temperature dependence of $^{63}(1/T_{1}T)_{c}$, the $^{63}$Cu nuclear spin-lattice relaxation rate $1/T_{1}$ divided by temperature $T$, measured in $B_{ext} = 9$~T applied along the c-axis.  We also show $^{63}(1/T_{1}T)_{a}$ and $^{63}(1/T_{2}T)_{a}$ measured with $B_{ext}~||~a$-axis, by normalizing their magnitude by multiplying a factor 1/3 and 1/15, respectively.  The red solid curve represents the best Curie-Weiss fit with $\theta_{CW} = 60$~K between $T_{c}=T_{spin} = 42$~K and $T_{mobility}\simeq 200$~K, which slightly underestimates the experimental results above $T_{mobility}$.  Also shown using the right axis is $^{139}(1/T_{1}T)$ measured at the $^{139}$La sites in $B_{ext} = 8.75$~T applied along the c-axis.
}\label{cs}
\end{figure}

In Fig.\ 1, we summarize $^{139}$La NMR lineshape  observed for the nuclear spin $I_{z}=+1/2$ to -1/2 central transition in an external magnetic field $B_{ext} = 8.7521$~T applied along the crystal c-axis.  At 295~K, the peak frequency is $^{139}f_{o} = 52.650$~MHz and the linewidth is as narrow as $\Delta f = 0.028$~MHz.  Using the nuclear gyromagnetic ratio $^{139}\gamma_{n} / 2\pi = 6.0146$~MHz/T, we estimate the apparent NMR frequency shift to be as small as $0.018 \pm 0.02$ \%, as expected for non-magnetic ions with closed shells and with a negligibly small second order nuclear quadrupole shift $\Delta \nu_{Q}^{(2)}$ by the electric field gradient (EFG). The latter implies that CuO$_6$ octahedra are not statically tilted away from the main principal axis of the EFG tensor along the c-axis above $T_{stage}$.  These $^{139}$La NMR results are also generic in the high temperature tetragonal (HTT) structure of La$_{2-x}$Sr$_{x}$CuO$_{4}$ \cite{Thurber2122, ImaiAspen, ArsenaultPRB2017}.

The $^{139}$La NMR lineshape gradually broadens below the stage-4 order at $T_{stage} \simeq 290$~K \cite{Wells_Science}.  In such a stage-4 structure, the elongated axis of the CuO$_6$ octahedra are statically tilted away from the crystal c-axis with a significant distribution in the tilting angle.  This tilting results in negative values of the second order nuclear quadrupole shift $\Delta \nu_{Q}^{(2)}$, and hence the line broadening.  The NMR lineshape remains fairly narrow down to $T_{mobility} \simeq 200$~K, presumably due to the motional narrowing effect caused by the slow motion of O$^{2-}$ ions.  As the excess O$^{2-}$ ions lose their mobility \cite{Kremer} and develop a superstructure within the plane \cite{Wells_Science, Xiong} near and below $T_{mobility} \simeq 200$~K, a broad, sawtooth shaped NMR line emerges underneath the narrow peak at $^{139}f_{o}$ with a distribution up to $\Delta \nu_{Q}^{(2)} \sim -0.5$~MHz.   

Ignoring the small asymmetry in the EFG tensor (because the first order quadrupole satellite peaks do not split when we apply $B_{ext}$ within the ab-plane), we may write $\Delta \nu_{Q}^{(2)} \simeq 15\nu_{Q}^{2}(1-cos^{2} \theta)(1-9cos^{2} \theta)/(16 \cdot ^{139}\gamma_{n} \cdot B_{ext}$), where the nuclear quadrupole frequency $\nu_{Q} = 5.55$~MHz as determined from the splitting between the first order quadrupolar satellite peaks at 45~K, and $\theta$ represents the angle between the external field $B_{ext}$ applied along the crystal c-axis and the main principal axis of the EFG tensor at the $^{139}$La  sites.  Thus the saw-tooth shaped NMR lineshape observed below $T_{mobility} \simeq 200$~K may be considered as a histogram of the static tilting angle $\theta$, as shown by the upper horizontal axis of Fig.\ 1.  We find that the most probable $\theta$ remains 0$^{\circ}$ but $\theta$ reaches $\theta \sim 25^{\circ}$ for the low frequency end of the spectrum.

Another important point to notice is that the needle-like, sharp central peak located at $^{139}f_{o}$ at 295~K remains visible on top of the broad, saw-tooth shaped lineshape down to $\sim 60$~K.  This means that, when averaged over the NMR measurement time scale, CuO$_6$ octahedra in some domains remain untilted far below $T_{stage}$ and $T_{mobility}$ down to $\sim 60$~K.  The half-width of the lineshape observed at the onset of charge order at 60~K, $\Delta f \sim 0.2$~MHz, is nearly identical with that of La$_{1.885}$Sr$_{0.115}$CuO$_{4}$ ($\sim 0.25$~MHz) observed at its charge order temperature $\sim 80$~K \cite{ArsenaultPRB2017}.  This means that the tilting angle distribution is comparable between La$_{2}$CuO$_{4+y}$ and La$_{1.885}$Sr$_{0.115}$CuO$_{4}$.  The $^{139}$La NMR lineshape in the latter, however, is nearly symmetrical with a Gaussian shape below $\sim 200$~K, and its peak is shifted from $^{139}f_{o}$.  The central value of the tilting angle is therefore non-zero in the case of La$_{1.885}$Sr$_{0.115}$CuO$_{4}$.

In Fig.\ 2, we summarize the temperature dependence of the nuclear spin-lattice relaxation rate $1/T_1$ divided by temperature $T$ measured at $^{139}f_{o}$.  The $T_{1}$ relaxation process showed a noticeable distribution due to the quadrupole relaxation effects induced by the motion of O$^{2-}$ ions as well as by the distribution of Cu spin fluctuation frequency scales.  Accordingly, we estimated  $1/T_1$  based on the force fit of the recovery data to the appropriate fitting function with multiple normal modes \cite{Narath}.  $1/T_1$  thus deduced generally captures the volume averaged behavior of spin dynamics of La$_2$CuO$_4$-based cuprates very well.  $^{139}(1/T_{1}T)$ exhibits a broad hump centered somewhat above $T_{mobility} \simeq 200$~K.  We attribute the enhancement of $1/T_1$ to the slow fluctuations of the EFG caused by the motion of O$^{2-}$ and dynamic distortion of the lattice.   The EFG contribution to $1/T_1$ takes the maximum value when the fluctuation frequency of the EFG slows down and becomes comparable to the NMR frequency.  $^{139}(1/T_{1}T)$ continues to decrease below $T_{mobility}$ as the fluctuation gradually diminishes.  Analogous enhancement of NMR relaxation rates are commonly observed in ionic conductors.  For example, the slow motion of Na$^{+}$ ions in the battery material  Na$_{1-x}$CoO$_2$ enhances $1/T_1$ at Na$^{+}$ sites dramatically near room temperature \cite{Weller}.  

\subsubsection{Low frequency Cu spin dynamics}      
In general, $1/T_{1}T$ probes the wave vector {\bf q} integral of the dynamical electron spin susceptibility at the NMR frequency $\chi''({\bf q}, f_{o})$.  $^{139}(1/T_{1}T)$ begins to grow rather sharply below $T_{charge}$.  Since a magnetic long range order is known to set in at $T_{spin} = 42$~K at the time scale of elastic neutron scattering \cite{LeePRB1999} and $\mu$SR measurements \cite{Savici}, our $^{139}(1/T_{1}T)$ results indicate that charge order triggers enhancement of low frequency Cu spin fluctuations, as noted before \cite{HuntPRL, HuntPRB}.  This enhancement is commonly observed for the charge ordered La$_2$CuO$_4$-based superconductors \cite{HuntPRL, ImaiAspen, HuntPRB, Mitrovic, BaekLaT1PRB2015, ImaiPRB2017, ArsenaultPRB2017}.  Inelastic neutron scattering measurements with low energy transfer also provide additional evidence for enhanced spin excitations below $T_{charge}$ in La$_2$CuO$_{4+y}$ \cite{LeePRB1999}, as well as in La$_{1.48}$Nd$_{0.4}$Sr$_{0.12}$CuO$_{4}$ \cite{TranquadaPRB59}, La$_{1.88}$Ba$_{0.12}$CuO$_{4}$ \cite{Fujita}, and La$_{1.885}$Sr$_{0.115}$CuO$_{4}$ \cite{RomerNeutron}.  Neutron scattering, however, probes only the bulk averaged behavior of spin excitations by integrating the signals from the entire volume of the samples.  We will come back to this point below.  

In Fig.\ 2, we also present $^{63}(1/T_{1}T)$ measured at the peak of the $^{63}$Cu NMR lineshape observed for the nuclear spin $I_{z} = +1/2$ to -1/2 central transition in Fig.\ 3(a) and (b). Since the magnetic hyperfine interaction of the $^{63}$Cu nuclear spin with Cu electron spins \cite{Tsuda, Mila-Rice} is nearly two orders of magnitude larger than that of $^{139}$La nuclear spins \cite{Nishihara}, the overall magnitude of $^{63}(1/T_{1}T)$ at $\sim 60$~K is greater by four orders of magnitude.   Accordingly, the contribution of the fluctuating EFG to the spin-lattice relaxation process is relatively small.  We observed only a small hump of  $^{63}(1/T_{1}T)$ near $T_{mobility}$.

Turning our attention to the low temperature region deep inside the stage-4 ordered state, we note that $^{63}(1/T_{1}T)$ increases smoothly down to $T_c$ obeying a Curie-Weiss law, which is indicative of the growth of antiferromagnetic spin correlations commonly observed in superconducting cuprates \cite{Imai1988, Barrett, MMP1990, Ohsugi1994, ImaiPRB2017}.  $^{63}(1/T_{1}T)$ begins to dive down at $T_c$ without exhibiting a Hebel-Slichter coherence peak.  This behavior is also prototypical of the d-wave superconductors \cite{Imai1988, Ishida}.  

Thus, the $^{63}(1/T_{1}T)$ results (except for a small hump near $T_{mobility}$) are representative of the canonical behavior expected for optimally doped high $T_c$ superconductors both above and below $T_c$ with no hints of anomalies associated with charge and spin orders. In contrast, $^{139}(1/T_{1}T)$ begins to increase below $T_{charge}$ and shows a maximum at $\sim33$~K.  This indicates that Cu spin fluctuations averaged over the entire volume of the sample are slowing down to the time scale of NMR measurements at $\sim33$~K.  How can we reconcile the apparent contradiction between $^{63}(1/T_{1}T)$ and $^{139}(1/T_{1}T)$ results?  

The key to understanding the dichotomy is that $^{63}(1/T_{1}T)$ and $^{139}(1/T_{1}T)$ do ${\it not}$ probe the same parts of the CuO$_2$ plane.  As noted earlier \cite{HuntPRL, SingerPRB, HuntPRB, ImaiPRB2017, ArsenaultPRB2017} and explained in detail in the next section, the $^{63}$Cu NMR signal intensity is gradually wiped out below $T_{charge}$, and hence $^{63}(1/T_{1}T)$ in Fig.\ 2 reflects only the behavior of Cu electrons  in a shrinking volume fraction of CuO$_2$ planes which has not been affected by charge order.  In contrast,  the $^{139}$La NMR signal intensity is conserved through $T_{charge}$ owing to the very small hyperfine interactions with Cu electron spins, and hence $^{139}(1/T_{1}T)$ probes the average behavior of the entire CuO$_2$ planes.  

\begin{figure}
\centering
\includegraphics[width=3.5in]{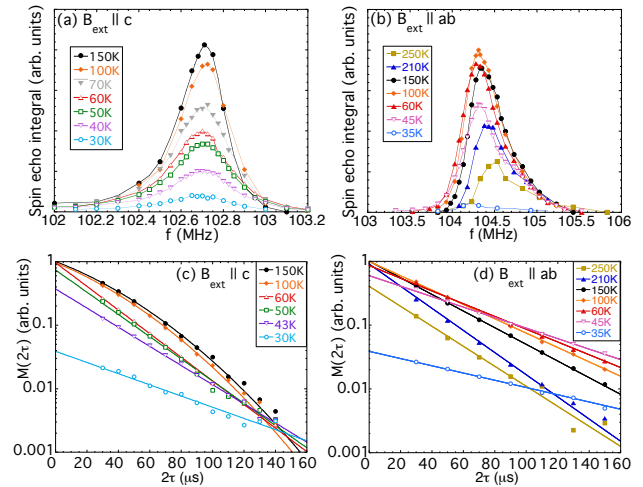}
\caption{Representative $^{63}$Cu NMR lineshapes measured in $B_{ext} = 9$~T applied along the crystal (a) c-axis and (b) a-axis at a constant delay time $\tau = 15~\mu$s.  Also shown in (c) and (d) are the spin echo decay curves in $B_{ext}~||~c$ and $B_{ext}~||~a$, respectively.  The magnitude of the spin echo is normalized at $\tau = 15~\mu$s using the integral of the lineshapes in (a) and (b). Solid lines are the best Gaussian-Lorentzian fit in (c) and Lorentzian fit in (d).  The extrapolation of the fit to $\tau = 0$ therefore yields the integrated intensity of the lineshape in the limit of $\tau = 0$.  For example, (c) shows that the intensity at $2\tau = 30~\mu$s observed at 60~K appears much smaller than at 100~K and 150~K; but the intensity in the limit of $2\tau = 0$ is conserved above $T_{charge}$.
}\label{cs}
\end{figure}

\begin{figure}
\centering
\includegraphics[width=3in]{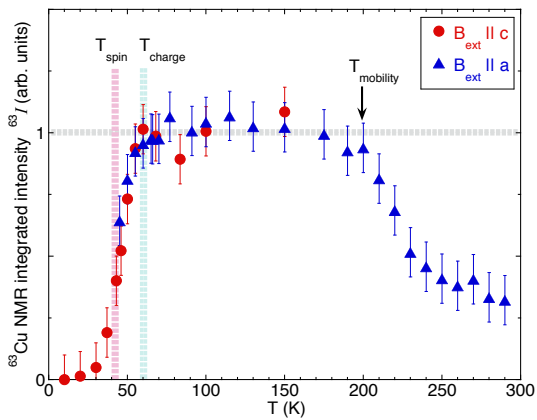}
\caption{Temperature dependence of the integrated intensity of $^{63}$Cu NMR lineshapes, $^{63}I$, corrected for the Boltzmann factor and the transverse relaxation effects based on the results in Fig.\ 3(c) and (d).  For $B_{ext}~||~a$, the large superconducting critical field makes the spin echo signal nearly unobservable due to the Meissner effects below $T_c$.   
}\label{cs}
\end{figure}

\subsubsection{$^{63}Cu$ NMR signal intensity anomaly below $T_{charge}$}      

In Fig.\ 3(a), we summarize the representative $^{63}$Cu NMR lineshapes observed for the $I_{z}=+1/2$ to -1/2 central transition in $B_{ext} = 9$~T applied along the c-axis.  We multiplied the spin echo intensity with temperature $T$ to account for the effect of the Boltzmann factor on the overall intensity.  The c-axis $^{63}$Cu NMR linewidth $^{63}\Delta f$ grows with decreasing temperature, similarly to the Curie-Weiss behavior of $^{63}\Delta f$ observed for La$_{1.885}$Sr$_{0.115}$CuO$_{4}$ \cite{ImaiPRB2017}.   The integrated intensity is proportional to the number of nuclear spins detected, and conserved between 150~K and 100~K.  

The integrated intensity of the lineshape appears to decrease progressively below 100~K down to $T_{charge} \simeq 60$~K, but it is simply because the transverse nuclear spin-spin relaxation becomes faster toward $T_{charge}$.  As shown in Fig.\ 3(c), the Gaussian curvature of the spin echo decay observed at 100~K and 150~K disappears, and the spin echo decay observed above $\tau \sim 10~\mu$s becomes Lorentzian (i.e. exponential) at $T_{charge}$ \cite{TouLBCOT2, HuntPRL}.  For this reason, when we measure the $^{63}$Cu NMR lineshape using the same pulse separation time $\tau = 15~\mu$s, the intensity appears smaller at $T_{charge}$ than at 100~K.  We can eliminate the influence of the transverse relaxation process on the integrated intensity, by extrapolating the spin echo decay curves in Fig.\ 3(c) to $\tau = 0$.  The integrated intensity $^{63}I$ in the limit of $\tau = 0$ maintains a constant value down to $T_{charge}$, as summarized in Fig.\ 4.

Below $T_{charge}$, however, the situation is very different.  Even if we extrapolate the spin echo decay to $\tau = 0$ and eliminate the transverse relaxation effect in the intensity, $^{63}I$ sharply decreases in the charge ordered state.  We confirmed that the integrated intensity of $^{139}$La NMR lineshape is conserved even below $T_{charge}$, and hence the reduction of $^{63}I$ below $T_{charge}$ down to $T_c$ is not an experimental artifact.    

In our recent comprehensive $^{63}$Cu NMR  study of  La$_{1.885}$Sr$_{0.115}$CuO$_{4}$, we used our state-of-the-art NMR spectrometer to investigate the NMR properties for a wide range of $\tau$ between 2~$\mu$s and 30~$\mu$s \cite{ImaiPRB2017}.  We demonstrated that $^{63}$Cu NMR properties show unprecedentedly strong dependence on $\tau$ below $T_{charge} \simeq 80$~K.  We also confirmed that the $^{63}$Cu NMR integrated intensity of the narrow central peak is gradually wiped out below  $T_{charge}$, because the missing spectral weight is transferred to a very broad, wing-like $^{63}$Cu NMR signal that appears symmetrically on both sides of the narrow central peak.  The wing-like signal from the charge ordered segments of the CuO$_2$ planes has very fast $T_1$ and $T_2$, and can be detected only with extremely short  $\tau \sim 2~\mu$s.  

Also summarized in Fig.\ 3(b) are the $^{63}$Cu NMR lineshapes observed in $B_{ext} = 9$~T applied along the crystal a-axis.  The spin echo decay is always Lorentzian in this field geometry as summarized in Fig.\ 4(d), because the Gaussian term induced by the indirect nuclear spin-spin coupling effect is motionally narrowed \cite{PenningtonPRB1989}.  We plot the temperature dependence of the Lorentzian $1/T_2$ divided by $T$ in Fig.\ 2 for comparison with $1/T_{1}T$.  Since the Redfield contribution $1/T_{2R}$ of the longitudinal relaxation process to the Lorentzian term is proportional to $1/T_{1}$ \cite{PenningtonPRB1989} and the contribution of the indirect spin-spin coupling term to $1/T_{2}T \propto \xi$ in the scaling limit, where $\xi$ is the spin-spin correlation length \cite{Sokol}, the observed temperature dependence of $1/T_{2}T$ is very similar to that of $1/T_{1}T$.  

We extrapolated the exponential spin echo decay to $\tau = 0$ in Fig.\ 3(d), and estimated the temperature dependence of $^{63}I$ for $B_{ext}~||~a$ in Fig.\ 4.   $^{63}I$ decreases above $T_{stage}$, because the hopping motion of oxygen ions wipe out the $^{63}$Cu NMR signals.  This is typical for ionic conductors.  In addition, $^{63}I$ decreases below $T_{charge}$, in agreement with the results for the $B_{ext}~||~c$ geometry.  In our recent measurements of La$_{1.885}$Sr$_{0.115}$CuO$_{4}$ with the $B_{ext}~||~a$ geometry (see Appendix B in \cite{ImaiPRB2017}), we were able to observe initial quick spin echo decay from $\tau = 2$ to 10~$\mu$s arising from the nuclear spins that appear as the wing-like signal for the $B_{ext}~||~c$ geometry.  It is their contribution that is missing below $T_{charge}$ in Fig.\ 4. 

Since the $B_{ext}~||~a$ configuration has one less fitting parameters in the extrapolation of the spin echo decay curves to $\tau = 0$, the $^{63}I$ data for $B_{ext}~||~a$ are less scattered than $B_{ext}~||~c$.  Moreover, the $B_{ext}~||~a$ geometry is free from the uncertainties in the extrapolation caused by the residual Gaussian curvature that manifests itself only for extremely short $\tau$ \cite{ImaiPRB2017}.  In fact, the onset of charge order $T_{charge} \simeq 60$~K in Fig.\ 4 agrees very well with the recent x-ray scattering reports \cite{He, Wells}. The downside of the $B_{ext}~||~a$ geometry is that the large superconducting critical field $B_{c2}$ makes NMR signal detection nearly impossible below $T_c$.   Overall, our results of $^{63}I$ below $T_{charge}$ are very similar to the case of La$_{1.885}$Sr$_{0.115}$CuO$_{4}$ \cite{ImaiPRB2017}, except that charge order in the present case takes place more quickly once it sets in.    

To understand the origin of the intensity anomaly in Fig.\ 4, it is useful to recall that $^{63}$Cu NMR signals in typical high $T_c$ cuprates has a large $1/T_{1} \sim 2 \times 10^{3}$~s$^{-1}$.  Enhancement of spin correlations by a factor of $\sim 5$ is sufficient to make $^{63}$Cu NMR spin echo signals nearly unobservable due to extremely fast relaxation rates, unless we resort to $\tau \sim 2~\mu$s using  modern NMR spectrometers.  That is why, for example, our earlier $^{63}$Cu NQR measurements in the paramagnetic state of La$_{2}$CuO$_{4}$ covered the temperature range only down to $\sim 400$~K, where $1/T_{1}$ reaches $ \sim 10^{4}$~s$^{-1}$ \cite{Imai1993_1, Imai1993_2}; the spin echo signal cannot be detected when $T_{1}$, and hence $T_{2}$, becomes so fast.  In fact, our recent $^{63}$Cu NMR study of La$_{1.885}$Sr$_{0.115}$CuO$_{4}$ showed that $1/T_{1}$ measured for the aforementioned wing-like NMR signal arising from charge ordered domains is $\sim 5$ times enhanced compared with the normally behaving residual $^{63}$Cu NMR signals below $T_{charge}$.  

Our consideration in the previous paragraph provides the key to reconciling the dichotomy exhibited by qualitatively different $^{63}(1/T_{1}T)$ and $^{139}(1/T_{1}T)$ in Fig.\  2.   The former reflects only the diminishing volume fraction of CuO$_2$ planes that has not been affected by charge order below $T_{charge}$.  In other words, our NMR results imply that charge order makes the CuO$_2$ planes extremely inhomogeneous in La$_{2}$CuO$_{4+y}$ as well as in La$_{1.885}$Sr$_{0.115}$CuO$_{4}$.  The charge ordered segments of CuO$_2$ planes have enhanced NMR relaxation rates, and they no longer contribute to the $^{63}$Cu NMR lineshapes nor $^{63}(1/T_{1}T)$ measured for the observable signals.  In contrast,  $^{139}$La NMR intensity is conserved except below $T_c$, where superconducting shielding effects limit the overall signal intensity.  Therefore, $^{139}(1/T_{1}T)$ probes the average behavior of the entire CuO$_2$ planes.  Likewise, the enhancement of the inelastic neutron scattering signals with low energy transfer below $T_{charge}$ probably originates from enhanced spin correlations in charge ordered domains. 

\section{Summary and Conclusions}
In this paper, we described the previously unpublished NMR data to shed a new light on the interplay between the stage-4 excess O$^{2-}$ order, charge order, spin order, and superconductivity in La$_{2}$CuO$_{4+y}$.   Our NMR results show that  La$_{2}$CuO$_{4+y}$ exhibit very peculiar NMR anomalies at $T_{charge} \simeq 60$~K, which we identified first at the charge order transition of the Nd co-doped La$_{1.48}$Nd$_{0.4}$Sr$_{0.12}$CuO$_{4}$ \cite{HuntPRL}.  Recent x-ray scattering experiments indeed detected the charge order Bragg peaks successfully below  $T_{charge} \simeq 60$~K.  We note that we had  pointed out the possibility of a charge order transition in La$_{2}$CuO$_{4+y}$ in \cite{LeePRB1999} based on the present NMR data.

We demonstrated that the structural distortion associated with the spatial order of the mobile O$^{2-}$ ions slow down to the NMR measurement time scale below $\sim 200$~K and alter the $^{139}$La NMR lineshape.  The static distortion of the lattice, however, does not complete down to $T_{charge}\simeq 60$~K.  Once charge order sets in, the transition reflected on NMR data is sharp, similarly to the case of charge order in La$_{1.48}$Nd$_{0.4}$Sr$_{0.12}$CuO$_{4}$ and La$_{1.88}$Ba$_{0.12}$CuO$_{4}$ that sets in immediately after the LTT structure is established \cite{HuntPRL, SingerPRB, HuntPRB}.  In contrast, charge order transition that takes place in the LTO structure of La$_{1.885}$Sr$_{0.115}$CuO$_{4}$ is much more gradual, although the onset temperature is as high as  $T_{charge} \simeq 80$~K.  It suggests that the static LTT structure is not essential for charge order in La$_{2}$CuO$_{4}$-based cuprates.  We recall that conventional charge density wave order in NbSe$_2$ nucleates near the defects at much higher temperatures than the bulk charge density wave transition \cite{STM}.   La$_{1.885}$Sr$_{0.115}$CuO$_{4}$ must be less disordered than La$_{1.48}$Nd$_{0.4}$Sr$_{0.12}$CuO$_{4}$ and possibly La$_{2}$CuO$_{4+y}$, and defects and/or domain boundaries in the cleaner CuO$_2$ planes may be playing analogous roles as those in NbSe$_2$.

$^{139}$La and $^{63}$Cu NMR properties in La$_{2}$CuO$_{4+y}$ observed near and below $T_{charge}$ are qualitatively the same as those previously observed for other charge ordered La$_{2}$CuO$_{4}$-based cuprates \cite{HuntPRL, SingerPRB, HuntPRB, ImaiPRB2017, ArsenaultPRB2017}.  We emphasize that charge order in this and other La$_{2}$CuO$_{4}$-based cuprates does not take place uniformly in space; the fact that $^{63}$Cu NMR signals from some segments are observable yet other segments become unobservable due to enhanced spin correlations in this and other La$_{2}$CuO$_{4}$-based cuprates implies that the charge ordered state is electronically very inhomogeneous.  Even far below $T_{charge}$, some segments of CuO$_2$ planes remain unaffected by charge order and subsequent spin order at $T_{spin}=42$~K.  The NMR properties in these domains are similar to those observed for optimally doped superconducting CuO$_2$ planes.   It is important to note that when the superconducting transition and spin order set in simultaneously at 42~K, $\sim 1/3$ of $^{63}$Cu NMR signals are still observable with properties very similar to those of optimally doped superconducting phase La$_{1.85}$Sr$_{0.15}$CuO$_{4}$.  We recently reported detailed characterization of the analogous two component behavior of the CuO$_2$ planes for La$_{1.885}$Sr$_{0.115}$CuO$_{4}$ \cite{ImaiPRB2017, ArsenaultPRB2017}, in which the volume fraction under the influence of enhanced spin correlations reach $\sim 100$~\% at $\sim 20$~K.  We refer readers to \cite{Kivelson_PNAS} for possible theoretical implications of these experimental findings, in particular the relation between the competing order and the volume fractions of the competing phases.

Unlike the typical second order magnetic phase transitions, $^{139}(1/T_{1}T)$ does not diverge at $T_{spin} = 42$~K, where elastic neutron scattering detect the onset of spin order.  This means that the spin fluctuations do not slow down to the NMR measurement time scale at $T_{spin} = 42$~K in most of the sample volume.  $^{139}(1/T_{1}T)$  exhibits a broad hump at $\sim 33$~K instead, signaling that on average, spin fluctuations slow down to the NMR frequency at a temperature much lower than the onset of spin order.  We note that $^{139}(1/T_{1}T)$ peaks at much lower temperature, $\sim 8$~K, in the case of La$_{1.885}$Sr$_{0.115}$CuO$_{4}$.  Thus both the charge and spin orders in the present case take place much more sharply than in La$_{1.885}$Sr$_{0.115}$CuO$_{4}$.  Finally, we recall that earlier $\mu$SR measurements detected the onset of static spin order at  $T_{spin} = 42$~K {\it in some volume} of the sample, but the volume fraction of the statically spin density wave ordered phase at the $\mu$SR time scale is less than $\sim 20$~\% even at $\sim 2$~K \cite{Savici}.  Thus our finding does not contradict with the $\mu$SR  results.  

\section{Acknowledgement}
T. I. thanks K.\ M.\ Shen and A.\ W. Hunt for their assistance in NMR data acquisition in the early stage of this work, and B.\ O.\ Wells and S. A. Kivelson for their  helpful communications during the course of preparing the manuscript.  All the NMR measurements were conducted at M.I.T. between 1997 and 1999 with the financial support by NSF DMR 96-23858 and NSF DMR 98-08941.  The work at McMaster was financially supported by NSERC and CIFAR.  The work at Stanford and SLAC was supported by the U.S. Department of Energy (DOE), Office of Science, Basic Energy Sciences, Materials Sciences and Engineering Division, under Contract No. DE-AC02-76SF00515.  



\begin{thebibliography}{49}%
\makeatletter
\providecommand \@ifxundefined [1]{%
 \@ifx{#1\undefined}
}%
\providecommand \@ifnum [1]{%
 \ifnum #1\expandafter \@firstoftwo
 \else \expandafter \@secondoftwo
 \fi
}%
\providecommand \@ifx [1]{%
 \ifx #1\expandafter \@firstoftwo
 \else \expandafter \@secondoftwo
 \fi
}%
\providecommand \natexlab [1]{#1}%
\providecommand \enquote  [1]{``#1''}%
\providecommand \bibnamefont  [1]{#1}%
\providecommand \bibfnamefont [1]{#1}%
\providecommand \citenamefont [1]{#1}%
\providecommand \href@noop [0]{\@secondoftwo}%
\providecommand \href [0]{\begingroup \@sanitize@url \@href}%
\providecommand \@href[1]{\@@startlink{#1}\@@href}%
\providecommand \@@href[1]{\endgroup#1\@@endlink}%
\providecommand \@sanitize@url [0]{\catcode `\\12\catcode `\$12\catcode
  `\&12\catcode `\#12\catcode `\^12\catcode `\_12\catcode `\%12\relax}%
\providecommand \@@startlink[1]{}%
\providecommand \@@endlink[0]{}%
\providecommand \url  [0]{\begingroup\@sanitize@url \@url }%
\providecommand \@url [1]{\endgroup\@href {#1}{\urlprefix }}%
\providecommand \urlprefix  [0]{URL }%
\providecommand \Eprint [0]{\href }%
\providecommand \doibase [0]{http://dx.doi.org/}%
\providecommand \selectlanguage [0]{\@gobble}%
\providecommand \bibinfo  [0]{\@secondoftwo}%
\providecommand \bibfield  [0]{\@secondoftwo}%
\providecommand \translation [1]{[#1]}%
\providecommand \BibitemOpen [0]{}%
\providecommand \bibitemStop [0]{}%
\providecommand \bibitemNoStop [0]{.\EOS\space}%
\providecommand \EOS [0]{\spacefactor3000\relax}%
\providecommand \BibitemShut  [1]{\csname bibitem#1\endcsname}%
\let\auto@bib@innerbib\@empty
\bibitem [{\citenamefont {Comin}\ and\ \citenamefont
  {Damascelli}(2016)}]{Comin_Review}%
  \BibitemOpen
  \bibfield  {author} {\bibinfo {author} {\bibfnamefont {R.}~\bibnamefont
  {Comin}}\ and\ \bibinfo {author} {\bibfnamefont {A.}~\bibnamefont
  {Damascelli}},\ }\href@noop {} {\bibfield  {journal} {\bibinfo  {journal}
  {Ann. Rev. of Cond. Matt. Phys.}\ }\textbf {\bibinfo {volume} {7}},\ \bibinfo
  {pages} {369} (\bibinfo {year} {2016})}\BibitemShut {NoStop}%
\bibitem [{\citenamefont {Tranquada}\ \emph {et~al.}(1995)\citenamefont
  {Tranquada}, \citenamefont {Sternlieb}, \citenamefont {Axe}, \citenamefont
  {Nakamura},\ and\ \citenamefont {Uchida}}]{Tranquada}%
  \BibitemOpen
  \bibfield  {author} {\bibinfo {author} {\bibfnamefont {J.~M.}\ \bibnamefont
  {Tranquada}}, \bibinfo {author} {\bibfnamefont {B.~J.}\ \bibnamefont
  {Sternlieb}}, \bibinfo {author} {\bibfnamefont {J.~D.}\ \bibnamefont {Axe}},
  \bibinfo {author} {\bibfnamefont {Y.}~\bibnamefont {Nakamura}}, \ and\
  \bibinfo {author} {\bibfnamefont {S.}~\bibnamefont {Uchida}},\ }\href@noop {}
  {\bibfield  {journal} {\bibinfo  {journal} {Nature}\ }\textbf {\bibinfo
  {volume} {375}},\ \bibinfo {pages} {561} (\bibinfo {year}
  {1995})}\BibitemShut {NoStop}%
\bibitem [{\citenamefont {Hunt}\ \emph {et~al.}(1999)\citenamefont {Hunt},
  \citenamefont {Singer}, \citenamefont {Thurber},\ and\ \citenamefont
  {Imai}}]{HuntPRL}%
  \BibitemOpen
  \bibfield  {author} {\bibinfo {author} {\bibfnamefont {A.~W.}\ \bibnamefont
  {Hunt}}, \bibinfo {author} {\bibfnamefont {P.~M.}\ \bibnamefont {Singer}},
  \bibinfo {author} {\bibfnamefont {K.~R.}\ \bibnamefont {Thurber}}, \ and\
  \bibinfo {author} {\bibfnamefont {T.}~\bibnamefont {Imai}},\ }\href {\doibase
  10.1103/PhysRevLett.82.4300} {\bibfield  {journal} {\bibinfo  {journal}
  {Phys. Rev. Lett.}\ }\textbf {\bibinfo {volume} {82}},\ \bibinfo {pages}
  {4300} (\bibinfo {year} {1999})}\BibitemShut {NoStop}%
\bibitem [{\citenamefont {Singer}\ \emph {et~al.}(1999)\citenamefont {Singer},
  \citenamefont {Hunt}, \citenamefont {Cederstr\"om},\ and\ \citenamefont
  {Imai}}]{SingerPRB}%
  \BibitemOpen
  \bibfield  {author} {\bibinfo {author} {\bibfnamefont {P.~M.}\ \bibnamefont
  {Singer}}, \bibinfo {author} {\bibfnamefont {A.~W.}\ \bibnamefont {Hunt}},
  \bibinfo {author} {\bibfnamefont {A.~F.}\ \bibnamefont {Cederstr\"om}}, \
  and\ \bibinfo {author} {\bibfnamefont {T.}~\bibnamefont {Imai}},\ }\href
  {\doibase 10.1103/PhysRevB.60.15345} {\bibfield  {journal} {\bibinfo
  {journal} {Phys. Rev. B}\ }\textbf {\bibinfo {volume} {60}},\ \bibinfo
  {pages} {15345} (\bibinfo {year} {1999})}\BibitemShut {NoStop}%
\bibitem [{\citenamefont {Hunt}\ \emph {et~al.}(2001)\citenamefont {Hunt},
  \citenamefont {Singer}, \citenamefont {Cederstr\"om},\ and\ \citenamefont
  {Imai}}]{HuntPRB}%
  \BibitemOpen
  \bibfield  {author} {\bibinfo {author} {\bibfnamefont {A.~W.}\ \bibnamefont
  {Hunt}}, \bibinfo {author} {\bibfnamefont {P.~M.}\ \bibnamefont {Singer}},
  \bibinfo {author} {\bibfnamefont {A.~F.}\ \bibnamefont {Cederstr\"om}}, \
  and\ \bibinfo {author} {\bibfnamefont {T.}~\bibnamefont {Imai}},\ }\href
  {\doibase 10.1103/PhysRevB.64.134525} {\bibfield  {journal} {\bibinfo
  {journal} {Phys. Rev. B}\ }\textbf {\bibinfo {volume} {64}},\ \bibinfo
  {pages} {134525} (\bibinfo {year} {2001})}\BibitemShut {NoStop}%
\bibitem [{\citenamefont {Service}(1999)}]{ServiceScience}%
  \BibitemOpen
  \bibfield  {author} {\bibinfo {author} {\bibfnamefont {R.}~\bibnamefont
  {Service}},\ }\href@noop {} {\bibfield  {journal} {\bibinfo  {journal}
  {Science}\ }\textbf {\bibinfo {volume} {283}},\ \bibinfo {pages} {1116 }
  (\bibinfo {year} {1999})}\BibitemShut {NoStop}%
\bibitem [{\citenamefont {Croft}\ \emph {et~al.}(2014)\citenamefont {Croft},
  \citenamefont {Lester}, \citenamefont {Senn}, \citenamefont {Bombardi},\ and\
  \citenamefont {Hayden}}]{Croft}%
  \BibitemOpen
  \bibfield  {author} {\bibinfo {author} {\bibfnamefont {T.~P.}\ \bibnamefont
  {Croft}}, \bibinfo {author} {\bibfnamefont {C.}~\bibnamefont {Lester}},
  \bibinfo {author} {\bibfnamefont {M.~S.}\ \bibnamefont {Senn}}, \bibinfo
  {author} {\bibfnamefont {A.}~\bibnamefont {Bombardi}}, \ and\ \bibinfo
  {author} {\bibfnamefont {S.~M.}\ \bibnamefont {Hayden}},\ }\href {\doibase
  10.1103/PhysRevB.89.224513} {\bibfield  {journal} {\bibinfo  {journal} {Phys.
  Rev. B}\ }\textbf {\bibinfo {volume} {89}},\ \bibinfo {pages} {224513}
  (\bibinfo {year} {2014})}\BibitemShut {NoStop}%
\bibitem [{\citenamefont {Thampy}\ \emph {et~al.}(2014)\citenamefont {Thampy},
  \citenamefont {Dean}, \citenamefont {Christensen}, \citenamefont {Steinke},
  \citenamefont {Islam}, \citenamefont {Oda}, \citenamefont {Ido},
  \citenamefont {Momono}, \citenamefont {Wilkins},\ and\ \citenamefont
  {Hill}}]{Thampy}%
  \BibitemOpen
  \bibfield  {author} {\bibinfo {author} {\bibfnamefont {V.}~\bibnamefont
  {Thampy}}, \bibinfo {author} {\bibfnamefont {M.~P.~M.}\ \bibnamefont {Dean}},
  \bibinfo {author} {\bibfnamefont {N.~B.}\ \bibnamefont {Christensen}},
  \bibinfo {author} {\bibfnamefont {L.}~\bibnamefont {Steinke}}, \bibinfo
  {author} {\bibfnamefont {Z.}~\bibnamefont {Islam}}, \bibinfo {author}
  {\bibfnamefont {M.}~\bibnamefont {Oda}}, \bibinfo {author} {\bibfnamefont
  {M.}~\bibnamefont {Ido}}, \bibinfo {author} {\bibfnamefont {N.}~\bibnamefont
  {Momono}}, \bibinfo {author} {\bibfnamefont {S.~B.}\ \bibnamefont {Wilkins}},
  \ and\ \bibinfo {author} {\bibfnamefont {J.~P.}\ \bibnamefont {Hill}},\
  }\href {\doibase 10.1103/PhysRevB.90.100510} {\bibfield  {journal} {\bibinfo
  {journal} {Phys. Rev. B}\ }\textbf {\bibinfo {volume} {90}},\ \bibinfo
  {pages} {100510} (\bibinfo {year} {2014})}\BibitemShut {NoStop}%
\bibitem [{\citenamefont {He}\ \emph {et~al.}(2017)\citenamefont {He},
  \citenamefont {Lee},\ and\ \citenamefont {Fujita}}]{He}%
  \BibitemOpen
  \bibfield  {author} {\bibinfo {author} {\bibfnamefont {W.}~\bibnamefont
  {He}}, \bibinfo {author} {\bibfnamefont {Y.~S.}\ \bibnamefont {Lee}}, \ and\
  \bibinfo {author} {\bibfnamefont {M.}~\bibnamefont {Fujita}},\ }\href@noop {}
  {} (\bibinfo {year} {2017}),\ \bibinfo {note} {unpublished}\BibitemShut
  {NoStop}%
\bibitem [{\citenamefont {Imai}\ \emph {et~al.}(2017)\citenamefont {Imai},
  \citenamefont {Takahashi}, \citenamefont {Arsenault}, \citenamefont {Acton},
  \citenamefont {Lee}, \citenamefont {He}, \citenamefont {Lee},\ and\
  \citenamefont {Fujita}}]{ImaiPRB2017}%
  \BibitemOpen
  \bibfield  {author} {\bibinfo {author} {\bibfnamefont {T.}~\bibnamefont
  {Imai}}, \bibinfo {author} {\bibfnamefont {S.~K.}\ \bibnamefont {Takahashi}},
  \bibinfo {author} {\bibfnamefont {A.}~\bibnamefont {Arsenault}}, \bibinfo
  {author} {\bibfnamefont {A.~W.}\ \bibnamefont {Acton}}, \bibinfo {author}
  {\bibfnamefont {D.}~\bibnamefont {Lee}}, \bibinfo {author} {\bibfnamefont
  {W.}~\bibnamefont {He}}, \bibinfo {author} {\bibfnamefont {Y.~S.}\
  \bibnamefont {Lee}}, \ and\ \bibinfo {author} {\bibfnamefont
  {M.}~\bibnamefont {Fujita}},\ }\href@noop {} {\bibfield  {journal} {\bibinfo
  {journal} {Phys. Rev. B}\ }\textbf {\bibinfo {volume} {96}},\ \bibinfo
  {pages} {224508} (\bibinfo {year} {2017})}\BibitemShut {NoStop}%
\bibitem [{\citenamefont {Arsenault}\ \emph {et~al.}(2018)\citenamefont
  {Arsenault}, \citenamefont {Takahashi}, \citenamefont {Imai}, \citenamefont
  {He}, \citenamefont {Lee},\ and\ \citenamefont {Fujita}}]{ArsenaultPRB2017}%
  \BibitemOpen
  \bibfield  {author} {\bibinfo {author} {\bibfnamefont {A.}~\bibnamefont
  {Arsenault}}, \bibinfo {author} {\bibfnamefont {S.~K.}\ \bibnamefont
  {Takahashi}}, \bibinfo {author} {\bibfnamefont {T.}~\bibnamefont {Imai}},
  \bibinfo {author} {\bibfnamefont {W.}~\bibnamefont {He}}, \bibinfo {author}
  {\bibfnamefont {Y.~S.}\ \bibnamefont {Lee}}, \ and\ \bibinfo {author}
  {\bibfnamefont {M.}~\bibnamefont {Fujita}},\ }\href {\doibase
  10.1103/PhysRevB.97.064511} {\bibfield  {journal} {\bibinfo  {journal} {Phys.
  Rev. B}\ }\textbf {\bibinfo {volume} {97}},\ \bibinfo {pages} {064511}
  (\bibinfo {year} {2018})}\BibitemShut {NoStop}%
\bibitem [{\citenamefont {Zhang~et al.}()}]{Wells}%
  \BibitemOpen
  \bibfield  {author} {\bibinfo {author} {\bibfnamefont {Z.}~\bibnamefont
  {Zhang~et al.}},\ }\href@noop {} {}\bibinfo {howpublished}
  {arXiv:1707.04367}\BibitemShut {NoStop}%
\bibitem [{\citenamefont {Lee}\ \emph {et~al.}(1999)\citenamefont {Lee},
  \citenamefont {Birgeneau}, \citenamefont {Kastner}, \citenamefont {Endoh},
  \citenamefont {Wakimoto}, \citenamefont {Yamada}, \citenamefont {Erwin},
  \citenamefont {Lee},\ and\ \citenamefont {Shirane}}]{LeePRB1999}%
  \BibitemOpen
  \bibfield  {author} {\bibinfo {author} {\bibfnamefont {Y.~S.}\ \bibnamefont
  {Lee}}, \bibinfo {author} {\bibfnamefont {R.~J.}\ \bibnamefont {Birgeneau}},
  \bibinfo {author} {\bibfnamefont {M.~A.}\ \bibnamefont {Kastner}}, \bibinfo
  {author} {\bibfnamefont {Y.}~\bibnamefont {Endoh}}, \bibinfo {author}
  {\bibfnamefont {S.}~\bibnamefont {Wakimoto}}, \bibinfo {author}
  {\bibfnamefont {K.}~\bibnamefont {Yamada}}, \bibinfo {author} {\bibfnamefont
  {R.~W.}\ \bibnamefont {Erwin}}, \bibinfo {author} {\bibfnamefont {S.-H.}\
  \bibnamefont {Lee}}, \ and\ \bibinfo {author} {\bibfnamefont
  {G.}~\bibnamefont {Shirane}},\ }\href {\doibase 10.1103/PhysRevB.60.3643}
  {\bibfield  {journal} {\bibinfo  {journal} {Phys. Rev. B}\ }\textbf {\bibinfo
  {volume} {60}},\ \bibinfo {pages} {3643} (\bibinfo {year}
  {1999})}\BibitemShut {NoStop}%
\bibitem [{\citenamefont {Savici}\ \emph {et~al.}(2002)\citenamefont {Savici},
  \citenamefont {Fudamoto}, \citenamefont {Gat}, \citenamefont {Ito},
  \citenamefont {Larkin}, \citenamefont {Uemura}, \citenamefont {Luke},
  \citenamefont {Kojima}, \citenamefont {Lee}, \citenamefont {Kastner},
  \citenamefont {Birgeneau},\ and\ \citenamefont {Yamada}}]{Savici}%
  \BibitemOpen
  \bibfield  {author} {\bibinfo {author} {\bibfnamefont {A.~T.}\ \bibnamefont
  {Savici}}, \bibinfo {author} {\bibfnamefont {Y.}~\bibnamefont {Fudamoto}},
  \bibinfo {author} {\bibfnamefont {I.~M.}\ \bibnamefont {Gat}}, \bibinfo
  {author} {\bibfnamefont {T.}~\bibnamefont {Ito}}, \bibinfo {author}
  {\bibfnamefont {M.~I.}\ \bibnamefont {Larkin}}, \bibinfo {author}
  {\bibfnamefont {Y.~J.}\ \bibnamefont {Uemura}}, \bibinfo {author}
  {\bibfnamefont {G.~M.}\ \bibnamefont {Luke}}, \bibinfo {author}
  {\bibfnamefont {K.~M.}\ \bibnamefont {Kojima}}, \bibinfo {author}
  {\bibfnamefont {Y.~S.}\ \bibnamefont {Lee}}, \bibinfo {author} {\bibfnamefont
  {M.~A.}\ \bibnamefont {Kastner}}, \bibinfo {author} {\bibfnamefont {R.~J.}\
  \bibnamefont {Birgeneau}}, \ and\ \bibinfo {author} {\bibfnamefont
  {K.}~\bibnamefont {Yamada}},\ }\href {\doibase 10.1103/PhysRevB.66.014524}
  {\bibfield  {journal} {\bibinfo  {journal} {Phys. Rev. B}\ }\textbf {\bibinfo
  {volume} {66}},\ \bibinfo {pages} {014524} (\bibinfo {year}
  {2002})}\BibitemShut {NoStop}%
\bibitem [{\citenamefont {Kimura}\ \emph {et~al.}(1999)\citenamefont {Kimura},
  \citenamefont {Hirota}, \citenamefont {Matsushita}, \citenamefont {Yamada},
  \citenamefont {Endoh}, \citenamefont {Lee}, \citenamefont {Majkrzak},
  \citenamefont {Erwin}, \citenamefont {Shirane}, \citenamefont {Greven},
  \citenamefont {Lee}, \citenamefont {Kastner},\ and\ \citenamefont
  {Birgeneau}}]{Kimura}%
  \BibitemOpen
  \bibfield  {author} {\bibinfo {author} {\bibfnamefont {H.}~\bibnamefont
  {Kimura}}, \bibinfo {author} {\bibfnamefont {K.}~\bibnamefont {Hirota}},
  \bibinfo {author} {\bibfnamefont {H.}~\bibnamefont {Matsushita}}, \bibinfo
  {author} {\bibfnamefont {K.}~\bibnamefont {Yamada}}, \bibinfo {author}
  {\bibfnamefont {Y.}~\bibnamefont {Endoh}}, \bibinfo {author} {\bibfnamefont
  {S.-H.}\ \bibnamefont {Lee}}, \bibinfo {author} {\bibfnamefont {C.~F.}\
  \bibnamefont {Majkrzak}}, \bibinfo {author} {\bibfnamefont {R.}~\bibnamefont
  {Erwin}}, \bibinfo {author} {\bibfnamefont {G.}~\bibnamefont {Shirane}},
  \bibinfo {author} {\bibfnamefont {M.}~\bibnamefont {Greven}}, \bibinfo
  {author} {\bibfnamefont {Y.~S.}\ \bibnamefont {Lee}}, \bibinfo {author}
  {\bibfnamefont {M.~A.}\ \bibnamefont {Kastner}}, \ and\ \bibinfo {author}
  {\bibfnamefont {R.~J.}\ \bibnamefont {Birgeneau}},\ }\href {\doibase
  10.1103/PhysRevB.59.6517} {\bibfield  {journal} {\bibinfo  {journal} {Phys.
  Rev. B}\ }\textbf {\bibinfo {volume} {59}},\ \bibinfo {pages} {6517}
  (\bibinfo {year} {1999})}\BibitemShut {NoStop}%
\bibitem [{\citenamefont {Jorgensen}\ \emph {et~al.}(1988)\citenamefont
  {Jorgensen}, \citenamefont {Dabrowski}, \citenamefont {Pei}, \citenamefont
  {Hinks}, \citenamefont {Soderholm}, \citenamefont {Morosin}, \citenamefont
  {Schirber}, \citenamefont {Venturini},\ and\ \citenamefont
  {Ginley}}]{Jorgensen}%
  \BibitemOpen
  \bibfield  {author} {\bibinfo {author} {\bibfnamefont {J.~D.}\ \bibnamefont
  {Jorgensen}}, \bibinfo {author} {\bibfnamefont {B.}~\bibnamefont
  {Dabrowski}}, \bibinfo {author} {\bibfnamefont {S.}~\bibnamefont {Pei}},
  \bibinfo {author} {\bibfnamefont {D.~G.}\ \bibnamefont {Hinks}}, \bibinfo
  {author} {\bibfnamefont {L.}~\bibnamefont {Soderholm}}, \bibinfo {author}
  {\bibfnamefont {B.}~\bibnamefont {Morosin}}, \bibinfo {author} {\bibfnamefont
  {J.~E.}\ \bibnamefont {Schirber}}, \bibinfo {author} {\bibfnamefont {E.~L.}\
  \bibnamefont {Venturini}}, \ and\ \bibinfo {author} {\bibfnamefont {D.~S.}\
  \bibnamefont {Ginley}},\ }\href {\doibase 10.1103/PhysRevB.38.11337}
  {\bibfield  {journal} {\bibinfo  {journal} {Phys. Rev. B}\ }\textbf {\bibinfo
  {volume} {38}},\ \bibinfo {pages} {11337} (\bibinfo {year}
  {1988})}\BibitemShut {NoStop}%
\bibitem [{\citenamefont {Chou}\ \emph {et~al.}(1992)\citenamefont {Chou},
  \citenamefont {Cho},\ and\ \citenamefont {Johnston}}]{Chou}%
  \BibitemOpen
  \bibfield  {author} {\bibinfo {author} {\bibfnamefont {F.~C.}\ \bibnamefont
  {Chou}}, \bibinfo {author} {\bibfnamefont {J.~H.}\ \bibnamefont {Cho}}, \
  and\ \bibinfo {author} {\bibfnamefont {D.~C.}\ \bibnamefont {Johnston}},\
  }\href@noop {} {\bibfield  {journal} {\bibinfo  {journal} {Physica C}\
  }\textbf {\bibinfo {volume} {197}},\ \bibinfo {pages} {303} (\bibinfo {year}
  {1992})}\BibitemShut {NoStop}%
\bibitem [{\citenamefont {Wells}\ \emph {et~al.}(1996)\citenamefont {Wells},
  \citenamefont {Birgeneau}, \citenamefont {Chou}, \citenamefont {Endoh},
  \citenamefont {Johnston}, \citenamefont {Kastner}, \citenamefont {Lee},
  \citenamefont {Shirane}, \citenamefont {Tranquada},\ and\ \citenamefont
  {Yamada}}]{Wells_ZP}%
  \BibitemOpen
  \bibfield  {author} {\bibinfo {author} {\bibfnamefont {B.}~\bibnamefont
  {Wells}}, \bibinfo {author} {\bibfnamefont {R.~J.}\ \bibnamefont
  {Birgeneau}}, \bibinfo {author} {\bibfnamefont {F.~C.}\ \bibnamefont {Chou}},
  \bibinfo {author} {\bibfnamefont {Y.}~\bibnamefont {Endoh}}, \bibinfo
  {author} {\bibfnamefont {D.~C.}\ \bibnamefont {Johnston}}, \bibinfo {author}
  {\bibfnamefont {M.~A.}\ \bibnamefont {Kastner}}, \bibinfo {author}
  {\bibfnamefont {Y.}~\bibnamefont {Lee}}, \bibinfo {author} {\bibfnamefont
  {G.}~\bibnamefont {Shirane}}, \bibinfo {author} {\bibfnamefont {J.~M.}\
  \bibnamefont {Tranquada}}, \ and\ \bibinfo {author} {\bibfnamefont
  {K.}~\bibnamefont {Yamada}},\ }\href@noop {} {\bibfield  {journal} {\bibinfo
  {journal} {Z. Physics B}\ }\textbf {\bibinfo {volume} {100}},\ \bibinfo
  {pages} {535} (\bibinfo {year} {1996})}\BibitemShut {NoStop}%
\bibitem [{\citenamefont {Wells}\ \emph {et~al.}(1997)\citenamefont {Wells},
  \citenamefont {Lee}, \citenamefont {Kastner}, \citenamefont {chrisianson},
  \citenamefont {Birgenau}, \citenamefont {Yamada}, \citenamefont {Endoh},\
  and\ \citenamefont {Shirane}}]{Wells_Science}%
  \BibitemOpen
  \bibfield  {author} {\bibinfo {author} {\bibfnamefont {B.~O.}\ \bibnamefont
  {Wells}}, \bibinfo {author} {\bibfnamefont {Y.~S.}\ \bibnamefont {Lee}},
  \bibinfo {author} {\bibfnamefont {M.~A.}\ \bibnamefont {Kastner}}, \bibinfo
  {author} {\bibfnamefont {R.~J.}\ \bibnamefont {chrisianson}}, \bibinfo
  {author} {\bibfnamefont {R.~J.}\ \bibnamefont {Birgenau}}, \bibinfo {author}
  {\bibfnamefont {K.}~\bibnamefont {Yamada}}, \bibinfo {author} {\bibfnamefont
  {Y.}~\bibnamefont {Endoh}}, \ and\ \bibinfo {author} {\bibfnamefont
  {G.}~\bibnamefont {Shirane}},\ }\href@noop {} {\bibfield  {journal} {\bibinfo
   {journal} {Science}\ }\textbf {\bibinfo {volume} {277}},\ \bibinfo {pages}
  {1067} (\bibinfo {year} {1997})}\BibitemShut {NoStop}%
\bibitem [{\citenamefont {Khaykovich}\ \emph {et~al.}(2002)\citenamefont
  {Khaykovich}, \citenamefont {Lee}, \citenamefont {Erwin}, \citenamefont
  {Lee}, \citenamefont {Wakimoto}, \citenamefont {Thomas}, \citenamefont
  {Kastner},\ and\ \citenamefont {Birgeneau}}]{KhaykovichPRB2002}%
  \BibitemOpen
  \bibfield  {author} {\bibinfo {author} {\bibfnamefont {B.}~\bibnamefont
  {Khaykovich}}, \bibinfo {author} {\bibfnamefont {Y.~S.}\ \bibnamefont {Lee}},
  \bibinfo {author} {\bibfnamefont {R.~W.}\ \bibnamefont {Erwin}}, \bibinfo
  {author} {\bibfnamefont {S.-H.}\ \bibnamefont {Lee}}, \bibinfo {author}
  {\bibfnamefont {S.}~\bibnamefont {Wakimoto}}, \bibinfo {author}
  {\bibfnamefont {K.~J.}\ \bibnamefont {Thomas}}, \bibinfo {author}
  {\bibfnamefont {M.~A.}\ \bibnamefont {Kastner}}, \ and\ \bibinfo {author}
  {\bibfnamefont {R.~J.}\ \bibnamefont {Birgeneau}},\ }\href {\doibase
  10.1103/PhysRevB.66.014528} {\bibfield  {journal} {\bibinfo  {journal} {Phys.
  Rev. B}\ }\textbf {\bibinfo {volume} {66}},\ \bibinfo {pages} {014528}
  (\bibinfo {year} {2002})}\BibitemShut {NoStop}%
\bibitem [{\citenamefont {Khaykovich}\ \emph {et~al.}(2003)\citenamefont
  {Khaykovich}, \citenamefont {Birgeneau}, \citenamefont {Chou}, \citenamefont
  {Erwin}, \citenamefont {Kastner}, \citenamefont {Lee}, \citenamefont {Lee},
  \citenamefont {Smeibidl}, \citenamefont {Vorderwisch},\ and\ \citenamefont
  {Wakimoto}}]{KhaykovichPRB2003}%
  \BibitemOpen
  \bibfield  {author} {\bibinfo {author} {\bibfnamefont {B.}~\bibnamefont
  {Khaykovich}}, \bibinfo {author} {\bibfnamefont {R.~J.}\ \bibnamefont
  {Birgeneau}}, \bibinfo {author} {\bibfnamefont {F.~C.}\ \bibnamefont {Chou}},
  \bibinfo {author} {\bibfnamefont {R.~W.}\ \bibnamefont {Erwin}}, \bibinfo
  {author} {\bibfnamefont {M.~A.}\ \bibnamefont {Kastner}}, \bibinfo {author}
  {\bibfnamefont {S.-H.}\ \bibnamefont {Lee}}, \bibinfo {author} {\bibfnamefont
  {Y.~S.}\ \bibnamefont {Lee}}, \bibinfo {author} {\bibfnamefont
  {P.}~\bibnamefont {Smeibidl}}, \bibinfo {author} {\bibfnamefont
  {P.}~\bibnamefont {Vorderwisch}}, \ and\ \bibinfo {author} {\bibfnamefont
  {S.}~\bibnamefont {Wakimoto}},\ }\href {\doibase 10.1103/PhysRevB.67.054501}
  {\bibfield  {journal} {\bibinfo  {journal} {Phys. Rev. B}\ }\textbf {\bibinfo
  {volume} {67}},\ \bibinfo {pages} {054501} (\bibinfo {year}
  {2003})}\BibitemShut {NoStop}%
\bibitem [{\citenamefont {Jorge}\ \emph {et~al.}(2004)\citenamefont {Jorge},
  \citenamefont {Jaime}, \citenamefont {Civale}, \citenamefont {Batista},
  \citenamefont {Zink}, \citenamefont {Hellman}, \citenamefont {Khaykovich},
  \citenamefont {Kastner}, \citenamefont {Lee},\ and\ \citenamefont
  {Birgeneau}}]{JorgePRB2004}%
  \BibitemOpen
  \bibfield  {author} {\bibinfo {author} {\bibfnamefont {G.~A.}\ \bibnamefont
  {Jorge}}, \bibinfo {author} {\bibfnamefont {M.}~\bibnamefont {Jaime}},
  \bibinfo {author} {\bibfnamefont {L.}~\bibnamefont {Civale}}, \bibinfo
  {author} {\bibfnamefont {C.~D.}\ \bibnamefont {Batista}}, \bibinfo {author}
  {\bibfnamefont {B.~L.}\ \bibnamefont {Zink}}, \bibinfo {author}
  {\bibfnamefont {F.}~\bibnamefont {Hellman}}, \bibinfo {author} {\bibfnamefont
  {B.}~\bibnamefont {Khaykovich}}, \bibinfo {author} {\bibfnamefont {M.~A.}\
  \bibnamefont {Kastner}}, \bibinfo {author} {\bibfnamefont {Y.~S.}\
  \bibnamefont {Lee}}, \ and\ \bibinfo {author} {\bibfnamefont {R.~J.}\
  \bibnamefont {Birgeneau}},\ }\href {\doibase 10.1103/PhysRevB.69.174506}
  {\bibfield  {journal} {\bibinfo  {journal} {Phys. Rev. B}\ }\textbf {\bibinfo
  {volume} {69}},\ \bibinfo {pages} {174506} (\bibinfo {year}
  {2004})}\BibitemShut {NoStop}%
\bibitem [{\citenamefont {Kremer}\ \emph {et~al.}(1993)\citenamefont {Kremer},
  \citenamefont {Hizhnyakov}, \citenamefont {Sigmund}, \citenamefont {Simon},\
  and\ \citenamefont {M\"{u}ller}}]{Kremer}%
  \BibitemOpen
  \bibfield  {author} {\bibinfo {author} {\bibfnamefont {R.~K.}\ \bibnamefont
  {Kremer}}, \bibinfo {author} {\bibfnamefont {V.}~\bibnamefont {Hizhnyakov}},
  \bibinfo {author} {\bibfnamefont {E.}~\bibnamefont {Sigmund}}, \bibinfo
  {author} {\bibfnamefont {A.}~\bibnamefont {Simon}}, \ and\ \bibinfo {author}
  {\bibfnamefont {K.~A.}\ \bibnamefont {M\"{u}ller}},\ }\href@noop {}
  {\bibfield  {journal} {\bibinfo  {journal} {Z. Physik B}\ }\textbf {\bibinfo
  {volume} {91}},\ \bibinfo {pages} {169} (\bibinfo {year} {1993})}\BibitemShut
  {NoStop}%
\bibitem [{\citenamefont {Fujita}\ \emph {et~al.}(2004)\citenamefont {Fujita},
  \citenamefont {Goka}, \citenamefont {Yamada}, \citenamefont {Tranquada},\
  and\ \citenamefont {Regnault}}]{Fujita}%
  \BibitemOpen
  \bibfield  {author} {\bibinfo {author} {\bibfnamefont {M.}~\bibnamefont
  {Fujita}}, \bibinfo {author} {\bibfnamefont {H.}~\bibnamefont {Goka}},
  \bibinfo {author} {\bibfnamefont {K.}~\bibnamefont {Yamada}}, \bibinfo
  {author} {\bibfnamefont {J.~M.}\ \bibnamefont {Tranquada}}, \ and\ \bibinfo
  {author} {\bibfnamefont {L.~P.}\ \bibnamefont {Regnault}},\ }\href {\doibase
  10.1103/PhysRevB.70.104517} {\bibfield  {journal} {\bibinfo  {journal} {Phys.
  Rev. B}\ }\textbf {\bibinfo {volume} {70}},\ \bibinfo {pages} {104517}
  (\bibinfo {year} {2004})}\BibitemShut {NoStop}%
\bibitem [{\citenamefont {Hammel}\ \emph {et~al.}(1993)\citenamefont {Hammel},
  \citenamefont {Reyes}, \citenamefont {Cheong}, \citenamefont {Fisk},\ and\
  \citenamefont {Schirber}}]{Hammel}%
  \BibitemOpen
  \bibfield  {author} {\bibinfo {author} {\bibfnamefont {P.~C.}\ \bibnamefont
  {Hammel}}, \bibinfo {author} {\bibfnamefont {A.~P.}\ \bibnamefont {Reyes}},
  \bibinfo {author} {\bibfnamefont {S.-W.}\ \bibnamefont {Cheong}}, \bibinfo
  {author} {\bibfnamefont {Z.}~\bibnamefont {Fisk}}, \ and\ \bibinfo {author}
  {\bibfnamefont {J.~E.}\ \bibnamefont {Schirber}},\ }\href {\doibase
  10.1103/PhysRevLett.71.440} {\bibfield  {journal} {\bibinfo  {journal} {Phys.
  Rev. Lett.}\ }\textbf {\bibinfo {volume} {71}},\ \bibinfo {pages} {440}
  (\bibinfo {year} {1993})}\BibitemShut {NoStop}%
\bibitem [{\citenamefont {Thurber}\ \emph {et~al.}(1997)\citenamefont
  {Thurber}, \citenamefont {Hunt}, \citenamefont {Imai}, \citenamefont {Chou},\
  and\ \citenamefont {Lee}}]{Thurber2122}%
  \BibitemOpen
  \bibfield  {author} {\bibinfo {author} {\bibfnamefont {K.~R.}\ \bibnamefont
  {Thurber}}, \bibinfo {author} {\bibfnamefont {A.~W.}\ \bibnamefont {Hunt}},
  \bibinfo {author} {\bibfnamefont {T.}~\bibnamefont {Imai}}, \bibinfo {author}
  {\bibfnamefont {F.~C.}\ \bibnamefont {Chou}}, \ and\ \bibinfo {author}
  {\bibfnamefont {Y.~S.}\ \bibnamefont {Lee}},\ }\href {\doibase
  10.1103/PhysRevLett.79.171} {\bibfield  {journal} {\bibinfo  {journal} {Phys.
  Rev. Lett.}\ }\textbf {\bibinfo {volume} {79}},\ \bibinfo {pages} {171}
  (\bibinfo {year} {1997})}\BibitemShut {NoStop}%
\bibitem [{\citenamefont {Imai}\ and\ \citenamefont {Hirota}()}]{ImaiAspen}%
  \BibitemOpen
  \bibfield  {author} {\bibinfo {author} {\bibfnamefont {T.}~\bibnamefont
  {Imai}}\ and\ \bibinfo {author} {\bibfnamefont {K.}~\bibnamefont {Hirota}},\
  }\href@noop {} {}\bibinfo {note} {Unpublished $^{63}$Cu, $^{139}$La, and
  $^{17}$O NMR work on a La$_{1.885}$Sr$_{0.115}$CuO$_{4}$ single crystal
  presented first at the Aspen Winter Conference on Quantum Criticality
  (January 1999)}\BibitemShut {NoStop}%
\bibitem [{\citenamefont {Xiong}\ \emph {et~al.}(1996)\citenamefont {Xiong},
  \citenamefont {Wochner}, \citenamefont {Moss}, \citenamefont {Cao},
  \citenamefont {Koga},\ and\ \citenamefont {Fujita}}]{Xiong}%
  \BibitemOpen
  \bibfield  {author} {\bibinfo {author} {\bibfnamefont {X.}~\bibnamefont
  {Xiong}}, \bibinfo {author} {\bibfnamefont {P.}~\bibnamefont {Wochner}},
  \bibinfo {author} {\bibfnamefont {S.~C.}\ \bibnamefont {Moss}}, \bibinfo
  {author} {\bibfnamefont {Y.}~\bibnamefont {Cao}}, \bibinfo {author}
  {\bibfnamefont {K.}~\bibnamefont {Koga}}, \ and\ \bibinfo {author}
  {\bibfnamefont {N.}~\bibnamefont {Fujita}},\ }\href {\doibase
  10.1103/PhysRevLett.76.2997} {\bibfield  {journal} {\bibinfo  {journal}
  {Phys. Rev. Lett.}\ }\textbf {\bibinfo {volume} {76}},\ \bibinfo {pages}
  {2997} (\bibinfo {year} {1996})}\BibitemShut {NoStop}%
\bibitem [{\citenamefont {Narath}(1967)}]{Narath}%
  \BibitemOpen
  \bibfield  {author} {\bibinfo {author} {\bibfnamefont {A.}~\bibnamefont
  {Narath}},\ }\href {\doibase 10.1103/PhysRev.162.320} {\bibfield  {journal}
  {\bibinfo  {journal} {Phys. Rev.}\ }\textbf {\bibinfo {volume} {162}},\
  \bibinfo {pages} {320} (\bibinfo {year} {1967})}\BibitemShut {NoStop}%
\bibitem [{\citenamefont {Weller}\ \emph {et~al.}(2009)\citenamefont {Weller},
  \citenamefont {Sacchetti}, \citenamefont {Ott}, \citenamefont
  {Mattenberger},\ and\ \citenamefont {Batlogg}}]{Weller}%
  \BibitemOpen
  \bibfield  {author} {\bibinfo {author} {\bibfnamefont {M.}~\bibnamefont
  {Weller}}, \bibinfo {author} {\bibfnamefont {A.}~\bibnamefont {Sacchetti}},
  \bibinfo {author} {\bibfnamefont {H.~R.}\ \bibnamefont {Ott}}, \bibinfo
  {author} {\bibfnamefont {K.}~\bibnamefont {Mattenberger}}, \ and\ \bibinfo
  {author} {\bibfnamefont {B.}~\bibnamefont {Batlogg}},\ }\href {\doibase
  10.1103/PhysRevLett.102.056401} {\bibfield  {journal} {\bibinfo  {journal}
  {Phys. Rev. Lett.}\ }\textbf {\bibinfo {volume} {102}},\ \bibinfo {pages}
  {056401} (\bibinfo {year} {2009})}\BibitemShut {NoStop}%
\bibitem [{\citenamefont {Mitrovi\ifmmode~\acute{c}\else \'{c}\fi{}}\ \emph
  {et~al.}(2008)\citenamefont {Mitrovi\ifmmode~\acute{c}\else \'{c}\fi{}},
  \citenamefont {Julien}, \citenamefont {de~Vaulx}, \citenamefont
  {Horvati\ifmmode~\acute{c}\else \'{c}\fi{}}, \citenamefont {Berthier},
  \citenamefont {Suzuki},\ and\ \citenamefont {Yamada}}]{Mitrovic}%
  \BibitemOpen
  \bibfield  {author} {\bibinfo {author} {\bibfnamefont {V.~F.}\ \bibnamefont
  {Mitrovi\ifmmode~\acute{c}\else \'{c}\fi{}}}, \bibinfo {author}
  {\bibfnamefont {M.-H.}\ \bibnamefont {Julien}}, \bibinfo {author}
  {\bibfnamefont {C.}~\bibnamefont {de~Vaulx}}, \bibinfo {author}
  {\bibfnamefont {M.}~\bibnamefont {Horvati\ifmmode~\acute{c}\else
  \'{c}\fi{}}}, \bibinfo {author} {\bibfnamefont {C.}~\bibnamefont {Berthier}},
  \bibinfo {author} {\bibfnamefont {T.}~\bibnamefont {Suzuki}}, \ and\ \bibinfo
  {author} {\bibfnamefont {K.}~\bibnamefont {Yamada}},\ }\href {\doibase
  10.1103/PhysRevB.78.014504} {\bibfield  {journal} {\bibinfo  {journal} {Phys.
  Rev. B}\ }\textbf {\bibinfo {volume} {78}},\ \bibinfo {pages} {014504}
  (\bibinfo {year} {2008})}\BibitemShut {NoStop}%
\bibitem [{\citenamefont {Baek}\ \emph {et~al.}(2015)\citenamefont {Baek},
  \citenamefont {Utz}, \citenamefont {H\"ucker}, \citenamefont {Gu},
  \citenamefont {B\"uchner},\ and\ \citenamefont {Grafe}}]{BaekLaT1PRB2015}%
  \BibitemOpen
  \bibfield  {author} {\bibinfo {author} {\bibfnamefont {S.-H.}\ \bibnamefont
  {Baek}}, \bibinfo {author} {\bibfnamefont {Y.}~\bibnamefont {Utz}}, \bibinfo
  {author} {\bibfnamefont {M.}~\bibnamefont {H\"ucker}}, \bibinfo {author}
  {\bibfnamefont {G.~D.}\ \bibnamefont {Gu}}, \bibinfo {author} {\bibfnamefont
  {B.}~\bibnamefont {B\"uchner}}, \ and\ \bibinfo {author} {\bibfnamefont
  {H.-J.}\ \bibnamefont {Grafe}},\ }\href {\doibase 10.1103/PhysRevB.92.155144}
  {\bibfield  {journal} {\bibinfo  {journal} {Phys. Rev. B}\ }\textbf {\bibinfo
  {volume} {92}},\ \bibinfo {pages} {155144} (\bibinfo {year}
  {2015})}\BibitemShut {NoStop}%
\bibitem [{\citenamefont {Tranquada}\ \emph {et~al.}(1999)\citenamefont
  {Tranquada}, \citenamefont {Ichikawa},\ and\ \citenamefont
  {Uchida}}]{TranquadaPRB59}%
  \BibitemOpen
  \bibfield  {author} {\bibinfo {author} {\bibfnamefont {J.~M.}\ \bibnamefont
  {Tranquada}}, \bibinfo {author} {\bibfnamefont {N.}~\bibnamefont {Ichikawa}},
  \ and\ \bibinfo {author} {\bibfnamefont {S.}~\bibnamefont {Uchida}},\ }\href
  {\doibase 10.1103/PhysRevB.59.14712} {\bibfield  {journal} {\bibinfo
  {journal} {Phys. Rev. B}\ }\textbf {\bibinfo {volume} {59}},\ \bibinfo
  {pages} {14712} (\bibinfo {year} {1999})}\BibitemShut {NoStop}%
\bibitem [{\citenamefont {R\o{}mer}\ \emph {et~al.}(2013)\citenamefont
  {R\o{}mer}, \citenamefont {Chang}, \citenamefont {Christensen}, \citenamefont
  {Andersen}, \citenamefont {Lefmann}, \citenamefont {M\"ahler}, \citenamefont
  {Gavilano}, \citenamefont {Gilardi}, \citenamefont {Niedermayer},
  \citenamefont {R\o{}nnow}, \citenamefont {Schneidewind}, \citenamefont
  {Link}, \citenamefont {Oda}, \citenamefont {Ido}, \citenamefont {Momono},\
  and\ \citenamefont {Mesot}}]{RomerNeutron}%
  \BibitemOpen
  \bibfield  {author} {\bibinfo {author} {\bibfnamefont {A.~T.}\ \bibnamefont
  {R\o{}mer}}, \bibinfo {author} {\bibfnamefont {J.}~\bibnamefont {Chang}},
  \bibinfo {author} {\bibfnamefont {N.~B.}\ \bibnamefont {Christensen}},
  \bibinfo {author} {\bibfnamefont {B.~M.}\ \bibnamefont {Andersen}}, \bibinfo
  {author} {\bibfnamefont {K.}~\bibnamefont {Lefmann}}, \bibinfo {author}
  {\bibfnamefont {L.}~\bibnamefont {M\"ahler}}, \bibinfo {author}
  {\bibfnamefont {J.}~\bibnamefont {Gavilano}}, \bibinfo {author}
  {\bibfnamefont {R.}~\bibnamefont {Gilardi}}, \bibinfo {author} {\bibfnamefont
  {C.}~\bibnamefont {Niedermayer}}, \bibinfo {author} {\bibfnamefont {H.~M.}\
  \bibnamefont {R\o{}nnow}}, \bibinfo {author} {\bibfnamefont {A.}~\bibnamefont
  {Schneidewind}}, \bibinfo {author} {\bibfnamefont {P.}~\bibnamefont {Link}},
  \bibinfo {author} {\bibfnamefont {M.}~\bibnamefont {Oda}}, \bibinfo {author}
  {\bibfnamefont {M.}~\bibnamefont {Ido}}, \bibinfo {author} {\bibfnamefont
  {N.}~\bibnamefont {Momono}}, \ and\ \bibinfo {author} {\bibfnamefont
  {J.}~\bibnamefont {Mesot}},\ }\href {\doibase 10.1103/PhysRevB.87.144513}
  {\bibfield  {journal} {\bibinfo  {journal} {Phys. Rev. B}\ }\textbf {\bibinfo
  {volume} {87}},\ \bibinfo {pages} {144513} (\bibinfo {year}
  {2013})}\BibitemShut {NoStop}%
\bibitem [{\citenamefont {Tsuda}\ \emph {et~al.}(1988)\citenamefont {Tsuda},
  \citenamefont {Shimizu}, \citenamefont {Yasuoka}, \citenamefont {Kishio},\
  and\ \citenamefont {Kitazawa}}]{Tsuda}%
  \BibitemOpen
  \bibfield  {author} {\bibinfo {author} {\bibfnamefont {T.}~\bibnamefont
  {Tsuda}}, \bibinfo {author} {\bibfnamefont {T.}~\bibnamefont {Shimizu}},
  \bibinfo {author} {\bibfnamefont {H.}~\bibnamefont {Yasuoka}}, \bibinfo
  {author} {\bibfnamefont {K.}~\bibnamefont {Kishio}}, \ and\ \bibinfo {author}
  {\bibfnamefont {K.}~\bibnamefont {Kitazawa}},\ }\href@noop {} {\bibfield
  {journal} {\bibinfo  {journal} {J. Phys. Soc. Jpn.}\ }\textbf {\bibinfo
  {volume} {57}},\ \bibinfo {pages} {2908} (\bibinfo {year}
  {1988})}\BibitemShut {NoStop}%
\bibitem [{\citenamefont {Mila}\ and\ \citenamefont {Rice}(1989)}]{Mila-Rice}%
  \BibitemOpen
  \bibfield  {author} {\bibinfo {author} {\bibfnamefont {F.}~\bibnamefont
  {Mila}}\ and\ \bibinfo {author} {\bibfnamefont {T.~M.}\ \bibnamefont
  {Rice}},\ }\href@noop {} {\bibfield  {journal} {\bibinfo  {journal} {Physica
  C}\ }\textbf {\bibinfo {volume} {157}},\ \bibinfo {pages} {561} (\bibinfo
  {year} {1989})}\BibitemShut {NoStop}%
\bibitem [{\citenamefont {Nishihara}\ \emph {et~al.}(1987)\citenamefont
  {Nishihara}, \citenamefont {Yasuoka}, \citenamefont {Shimizu}, \citenamefont
  {Tsuda}, \citenamefont {Imai}, \citenamefont {Sasaki}, \citenamefont {Kanbe},
  \citenamefont {Kishio}, \citenamefont {Kitazawa},\ and\ \citenamefont
  {Fueki}}]{Nishihara}%
  \BibitemOpen
  \bibfield  {author} {\bibinfo {author} {\bibfnamefont {H.}~\bibnamefont
  {Nishihara}}, \bibinfo {author} {\bibfnamefont {H.}~\bibnamefont {Yasuoka}},
  \bibinfo {author} {\bibfnamefont {T.}~\bibnamefont {Shimizu}}, \bibinfo
  {author} {\bibfnamefont {T.}~\bibnamefont {Tsuda}}, \bibinfo {author}
  {\bibfnamefont {T.}~\bibnamefont {Imai}}, \bibinfo {author} {\bibfnamefont
  {S.}~\bibnamefont {Sasaki}}, \bibinfo {author} {\bibfnamefont
  {S.}~\bibnamefont {Kanbe}}, \bibinfo {author} {\bibfnamefont
  {K.}~\bibnamefont {Kishio}}, \bibinfo {author} {\bibfnamefont
  {K.}~\bibnamefont {Kitazawa}}, \ and\ \bibinfo {author} {\bibfnamefont
  {K.}~\bibnamefont {Fueki}},\ }\href@noop {} {\bibfield  {journal} {\bibinfo
  {journal} {J. Phys. Soc. Jpn.}\ }\textbf {\bibinfo {volume} {56}},\ \bibinfo
  {pages} {4559} (\bibinfo {year} {1987})}\BibitemShut {NoStop}%
\bibitem [{\citenamefont {Imai}\ \emph {et~al.}(1988)\citenamefont {Imai},
  \citenamefont {Shimizu}, \citenamefont {Yasuoka}, \citenamefont {Ueda},\ and\
  \citenamefont {Kosuge}}]{Imai1988}%
  \BibitemOpen
  \bibfield  {author} {\bibinfo {author} {\bibfnamefont {T.}~\bibnamefont
  {Imai}}, \bibinfo {author} {\bibfnamefont {T.}~\bibnamefont {Shimizu}},
  \bibinfo {author} {\bibfnamefont {H.}~\bibnamefont {Yasuoka}}, \bibinfo
  {author} {\bibfnamefont {Y.}~\bibnamefont {Ueda}}, \ and\ \bibinfo {author}
  {\bibfnamefont {K.}~\bibnamefont {Kosuge}},\ }\href@noop {} {\bibfield
  {journal} {\bibinfo  {journal} {J. Phys. Soc. Jpn.}\ }\textbf {\bibinfo
  {volume} {57}},\ \bibinfo {pages} {2280} (\bibinfo {year}
  {1988})}\BibitemShut {NoStop}%
\bibitem [{\citenamefont {Barrett}\ \emph {et~al.}(1990)\citenamefont
  {Barrett}, \citenamefont {Durand}, \citenamefont {Pennington}, \citenamefont
  {Slichter}, \citenamefont {Friedmann}, \citenamefont {Rice},\ and\
  \citenamefont {Ginsberg}}]{Barrett}%
  \BibitemOpen
  \bibfield  {author} {\bibinfo {author} {\bibfnamefont {S.~E.}\ \bibnamefont
  {Barrett}}, \bibinfo {author} {\bibfnamefont {D.~J.}\ \bibnamefont {Durand}},
  \bibinfo {author} {\bibfnamefont {C.~H.}\ \bibnamefont {Pennington}},
  \bibinfo {author} {\bibfnamefont {C.~P.}\ \bibnamefont {Slichter}}, \bibinfo
  {author} {\bibfnamefont {T.~A.}\ \bibnamefont {Friedmann}}, \bibinfo {author}
  {\bibfnamefont {J.~P.}\ \bibnamefont {Rice}}, \ and\ \bibinfo {author}
  {\bibfnamefont {D.~M.}\ \bibnamefont {Ginsberg}},\ }\href {\doibase
  10.1103/PhysRevB.41.6283} {\bibfield  {journal} {\bibinfo  {journal} {Phys.
  Rev. B}\ }\textbf {\bibinfo {volume} {41}},\ \bibinfo {pages} {6283}
  (\bibinfo {year} {1990})}\BibitemShut {NoStop}%
\bibitem [{\citenamefont {Millis}\ \emph {et~al.}(1990)\citenamefont {Millis},
  \citenamefont {Monien},\ and\ \citenamefont {Pines}}]{MMP1990}%
  \BibitemOpen
  \bibfield  {author} {\bibinfo {author} {\bibfnamefont {A.~J.}\ \bibnamefont
  {Millis}}, \bibinfo {author} {\bibfnamefont {H.}~\bibnamefont {Monien}}, \
  and\ \bibinfo {author} {\bibfnamefont {D.}~\bibnamefont {Pines}},\ }\href
  {\doibase 10.1103/PhysRevB.42.167} {\bibfield  {journal} {\bibinfo  {journal}
  {Phys. Rev. B}\ }\textbf {\bibinfo {volume} {42}},\ \bibinfo {pages} {167}
  (\bibinfo {year} {1990})}\BibitemShut {NoStop}%
\bibitem [{\citenamefont {Ohsugi}\ \emph {et~al.}(1994)\citenamefont {Ohsugi},
  \citenamefont {Kitaoka}, \citenamefont {Ishida}, \citenamefont {G.-q.},\ and\
  \citenamefont {K.}}]{Ohsugi1994}%
  \BibitemOpen
  \bibfield  {author} {\bibinfo {author} {\bibfnamefont {S.}~\bibnamefont
  {Ohsugi}}, \bibinfo {author} {\bibfnamefont {Y.}~\bibnamefont {Kitaoka}},
  \bibinfo {author} {\bibfnamefont {K.}~\bibnamefont {Ishida}}, \bibinfo
  {author} {\bibfnamefont {Z.}~\bibnamefont {G.-q.}}, \ and\ \bibinfo {author}
  {\bibfnamefont {A.}~\bibnamefont {K.}},\ }\href@noop {} {\bibfield  {journal}
  {\bibinfo  {journal} {J. Phys. Soc. Jpn.}\ }\textbf {\bibinfo {volume}
  {63}},\ \bibinfo {pages} {700} (\bibinfo {year} {1994})}\BibitemShut
  {NoStop}%
\bibitem [{\citenamefont {Ishida}\ \emph {et~al.}(1989)\citenamefont {Ishida},
  \citenamefont {Kitaoka},\ and\ \citenamefont {Asayama}}]{Ishida}%
  \BibitemOpen
  \bibfield  {author} {\bibinfo {author} {\bibfnamefont {K.}~\bibnamefont
  {Ishida}}, \bibinfo {author} {\bibfnamefont {Y.}~\bibnamefont {Kitaoka}}, \
  and\ \bibinfo {author} {\bibfnamefont {K.}~\bibnamefont {Asayama}},\
  }\href@noop {} {\bibfield  {journal} {\bibinfo  {journal} {J. Phys. Soc.
  Jpn.}\ }\textbf {\bibinfo {volume} {58}},\ \bibinfo {pages} {36} (\bibinfo
  {year} {1989})}\BibitemShut {NoStop}%
\bibitem [{\citenamefont {Tou}\ \emph {et~al.}(1992)\citenamefont {Tou},
  \citenamefont {Matsumura},\ and\ \citenamefont {Yamagata}}]{TouLBCOT2}%
  \BibitemOpen
  \bibfield  {author} {\bibinfo {author} {\bibfnamefont {H.}~\bibnamefont
  {Tou}}, \bibinfo {author} {\bibfnamefont {M.}~\bibnamefont {Matsumura}}, \
  and\ \bibinfo {author} {\bibfnamefont {H.}~\bibnamefont {Yamagata}},\
  }\href@noop {} {\bibfield  {journal} {\bibinfo  {journal} {J. Phys. Soc.
  Jpn.}\ }\textbf {\bibinfo {volume} {61}},\ \bibinfo {pages} {1477} (\bibinfo
  {year} {1992})}\BibitemShut {NoStop}%
\bibitem [{\citenamefont {Pennington}\ \emph {et~al.}(1989)\citenamefont
  {Pennington}, \citenamefont {Durand}, \citenamefont {Slichter}, \citenamefont
  {Rice}, \citenamefont {Bukowski},\ and\ \citenamefont
  {Ginsberg}}]{PenningtonPRB1989}%
  \BibitemOpen
  \bibfield  {author} {\bibinfo {author} {\bibfnamefont {C.~H.}\ \bibnamefont
  {Pennington}}, \bibinfo {author} {\bibfnamefont {D.~J.}\ \bibnamefont
  {Durand}}, \bibinfo {author} {\bibfnamefont {C.~P.}\ \bibnamefont
  {Slichter}}, \bibinfo {author} {\bibfnamefont {J.~P.}\ \bibnamefont {Rice}},
  \bibinfo {author} {\bibfnamefont {E.~D.}\ \bibnamefont {Bukowski}}, \ and\
  \bibinfo {author} {\bibfnamefont {D.~M.}\ \bibnamefont {Ginsberg}},\ }\href
  {\doibase 10.1103/PhysRevB.39.274} {\bibfield  {journal} {\bibinfo  {journal}
  {Phys. Rev. B}\ }\textbf {\bibinfo {volume} {39}},\ \bibinfo {pages} {274}
  (\bibinfo {year} {1989})}\BibitemShut {NoStop}%
\bibitem [{\citenamefont {Sokol}\ and\ \citenamefont {Pines}(1993)}]{Sokol}%
  \BibitemOpen
  \bibfield  {author} {\bibinfo {author} {\bibfnamefont {A.}~\bibnamefont
  {Sokol}}\ and\ \bibinfo {author} {\bibfnamefont {D.}~\bibnamefont {Pines}},\
  }\href {\doibase 10.1103/PhysRevLett.71.2813} {\bibfield  {journal} {\bibinfo
   {journal} {Phys. Rev. Lett.}\ }\textbf {\bibinfo {volume} {71}},\ \bibinfo
  {pages} {2813} (\bibinfo {year} {1993})}\BibitemShut {NoStop}%
\bibitem [{\citenamefont {Imai}\ \emph
  {et~al.}(1993{\natexlab{a}})\citenamefont {Imai}, \citenamefont {Slichter},
  \citenamefont {Yoshimura},\ and\ \citenamefont {Kosuge}}]{Imai1993_1}%
  \BibitemOpen
  \bibfield  {author} {\bibinfo {author} {\bibfnamefont {T.}~\bibnamefont
  {Imai}}, \bibinfo {author} {\bibfnamefont {C.~P.}\ \bibnamefont {Slichter}},
  \bibinfo {author} {\bibfnamefont {K.}~\bibnamefont {Yoshimura}}, \ and\
  \bibinfo {author} {\bibfnamefont {K.}~\bibnamefont {Kosuge}},\ }\href
  {\doibase 10.1103/PhysRevLett.70.1002} {\bibfield  {journal} {\bibinfo
  {journal} {Phys. Rev. Lett.}\ }\textbf {\bibinfo {volume} {70}},\ \bibinfo
  {pages} {1002} (\bibinfo {year} {1993}{\natexlab{a}})}\BibitemShut {NoStop}%
\bibitem [{\citenamefont {Imai}\ \emph
  {et~al.}(1993{\natexlab{b}})\citenamefont {Imai}, \citenamefont {Slichter},
  \citenamefont {Yoshimura}, \citenamefont {Katoh},\ and\ \citenamefont
  {Kosuge}}]{Imai1993_2}%
  \BibitemOpen
  \bibfield  {author} {\bibinfo {author} {\bibfnamefont {T.}~\bibnamefont
  {Imai}}, \bibinfo {author} {\bibfnamefont {C.~P.}\ \bibnamefont {Slichter}},
  \bibinfo {author} {\bibfnamefont {K.}~\bibnamefont {Yoshimura}}, \bibinfo
  {author} {\bibfnamefont {M.}~\bibnamefont {Katoh}}, \ and\ \bibinfo {author}
  {\bibfnamefont {K.}~\bibnamefont {Kosuge}},\ }\href {\doibase
  10.1103/PhysRevLett.71.1254} {\bibfield  {journal} {\bibinfo  {journal}
  {Phys. Rev. Lett.}\ }\textbf {\bibinfo {volume} {71}},\ \bibinfo {pages}
  {1254} (\bibinfo {year} {1993}{\natexlab{b}})}\BibitemShut {NoStop}%
\bibitem [{\citenamefont {Arguello}\ \emph {et~al.}(2014)\citenamefont
  {Arguello}, \citenamefont {Chockalingam}, \citenamefont {Rosenthal},
  \citenamefont {Zhao}, \citenamefont {Guti\'errez}, \citenamefont {Kang},
  \citenamefont {Chung}, \citenamefont {Fernandes}, \citenamefont {Jia},
  \citenamefont {Millis}, \citenamefont {Cava},\ and\ \citenamefont
  {Pasupathy}}]{STM}%
  \BibitemOpen
  \bibfield  {author} {\bibinfo {author} {\bibfnamefont {C.~J.}\ \bibnamefont
  {Arguello}}, \bibinfo {author} {\bibfnamefont {S.~P.}\ \bibnamefont
  {Chockalingam}}, \bibinfo {author} {\bibfnamefont {E.~P.}\ \bibnamefont
  {Rosenthal}}, \bibinfo {author} {\bibfnamefont {L.}~\bibnamefont {Zhao}},
  \bibinfo {author} {\bibfnamefont {C.}~\bibnamefont {Guti\'errez}}, \bibinfo
  {author} {\bibfnamefont {J.~H.}\ \bibnamefont {Kang}}, \bibinfo {author}
  {\bibfnamefont {W.~C.}\ \bibnamefont {Chung}}, \bibinfo {author}
  {\bibfnamefont {R.~M.}\ \bibnamefont {Fernandes}}, \bibinfo {author}
  {\bibfnamefont {S.}~\bibnamefont {Jia}}, \bibinfo {author} {\bibfnamefont
  {A.~J.}\ \bibnamefont {Millis}}, \bibinfo {author} {\bibfnamefont {R.~J.}\
  \bibnamefont {Cava}}, \ and\ \bibinfo {author} {\bibfnamefont {A.~N.}\
  \bibnamefont {Pasupathy}},\ }\href {\doibase 10.1103/PhysRevB.89.235115}
  {\bibfield  {journal} {\bibinfo  {journal} {Phys. Rev. B}\ }\textbf {\bibinfo
  {volume} {89}},\ \bibinfo {pages} {235115} (\bibinfo {year}
  {2014})}\BibitemShut {NoStop}%
\bibitem [{\citenamefont {Kivelson}\ \emph {et~al.}(2001)\citenamefont
  {Kivelson}, \citenamefont {Aepplie},\ and\ \citenamefont
  {Emery}}]{Kivelson_PNAS}%
  \BibitemOpen
  \bibfield  {author} {\bibinfo {author} {\bibfnamefont {S.~A.}\ \bibnamefont
  {Kivelson}}, \bibinfo {author} {\bibfnamefont {G.}~\bibnamefont {Aepplie}}, \
  and\ \bibinfo {author} {\bibfnamefont {V.~J.}\ \bibnamefont {Emery}},\
  }\href@noop {} {\bibfield  {journal} {\bibinfo  {journal} {Proc. Nat. Acad.
  Sci.}\ }\textbf {\bibinfo {volume} {98}},\ \bibinfo {pages} {11903} (\bibinfo
  {year} {2001})}\BibitemShut {NoStop}%
\end{thebibliography}

%

\end{document}